\documentclass[aip, amsmath, amssymb, reprint]{revtex4-2}
\usepackage{graphicx,verbatim}
\usepackage{braket}
\usepackage{color}
\usepackage{tikz}
\usetikzlibrary{positioning,calc}
\usepackage[compat=1.1.0]{tikz-feynman} 

\usepackage{soul}
\usepackage[normalem]{ulem}
\usepackage[x11names]{xcolor}

\usepackage{placeins}
\usepackage{float}

\usepackage[
    colorlinks=true,
    linkcolor=RoyalBlue4,
    citecolor=RoyalBlue4,
    urlcolor=RoyalBlue4,
    breaklinks=true
]{hyperref}

\begin{document}

\title{Validity and Limits of Low Order Hybridization Expansion Approaches for Multi-Orbital Systems}

\author{Dolev Goldberger}
 \affiliation{School of Physics, Tel Aviv University, Tel Aviv 6997801, Israel}
\author{Ido Zemach}
 \affiliation{School of Physics, Tel Aviv University, Tel Aviv 6997801, Israel}
\author{Lei Zhang}
 \affiliation{Department of Physics, University of Michigan, Ann Arbor, Michigan 48109, USA}
\author{Yang Yu}
 \affiliation{Department of Physics, University of Michigan, Ann Arbor, Michigan 48109, USA}
\author{Emanuel Gull}
 \email{egull@umich.edu}
 \affiliation{Department of Physics, University of Michigan, Ann Arbor, Michigan 48109, USA}
 \affiliation{Institute of Theoretical Physics, Faculty of Physics, University of Warsaw, Warsaw, Poland}
\author{Guy Cohen}
\email{gcohen@tau.ac.il}
 \affiliation{The Raymond and Beverley Sackler Center for Computational Molecular and Materials Science, Tel Aviv University, Tel Aviv 6997801, Israel}
 \affiliation{School of Chemistry, Tel Aviv University, Tel Aviv 6997801, Israel}
\author{Andr\'e Erpenbeck}
\email{andre.erpenbeck@uga.edu}
 \affiliation{Department of Physics and Astronomy, University of Georgia, Athens, Georgia 30602, USA}
 \affiliation{Center for Simulational Physics, University of Georgia, Athens, Georgia 30602, USA}

\date{\today}

\begin{abstract}
    Low-order hybridization expansion methods such as the non-crossing approximation (NCA) and the one-crossing approximation (OCA) are widely used impurity solvers in the study of strongly correlated systems, yet their accuracy in genuine multi-orbital settings remains poorly understood.
    Using the decoupled orbital limit as a controlled reference point, we derive analytic results connecting multi-orbital restricted propagators and Green's functions to their single-orbital counterparts, identify the diagrammatic mechanisms responsible for the breakdown of low-order methods in multi-orbital settings, and determine their regimes of applicability.
    Our central finding is that the accuracy of these methods is governed by the least correlated orbital: i.e., the orbital with the most rapidly decaying retarded Green’s function.
    That orbital's properties are transferred to all other orbitals through a spurious coupling generated by the truncated expansion, thereby suppressing correlation-induced features such as the Kondo resonance.
    This occurs even in orbitals that are themselves strongly correlated within single-orbital calculations using the same approximation scheme.
    We confirm this numerically across representative two-orbital model systems in the steady-state, systematically identifying the parameter regimes in which low-order methods succeed or fail.
    Our results provide a practical guide for assessing when insights from single-orbital calculations carry over to multi-orbital settings, and serve as a benchmark for the development and validation of higher-order multi-orbital impurity solvers.
\end{abstract}

\maketitle

\section{Introduction}
    Quantum impurity models are systems where a small, interacting region is coupled to one or more infinite noninteracting baths.
    Their dynamics plays a central role in a variety of fields.
    For example, models of this type are used to describe transport through mesoscopic quantum dots\cite{datta_quantum_2005,haug_quantum_2008} and molecular electronics;\cite{nitzan_electron_2003,tao_electron_2006,selzer_single-molecule_2006} solvation dynamics;\cite{nitzan_chemical_2006,bagchi_solvation_2010} and, by means of the dynamical mean field theory (DMFT),\cite{Georges_Dynamical_1996} the electronic structure of strongly correlated materials.
    As these applications evolve, the need for reliable simulations of increasingly general models grows with them.

    An important class of diagrammatic methods relevant to quantum impurity models is often referred to as ``non-crossing approximations'' (NCA) because they sum certain subsets of Feynman diagrams where lines do not cross.
    Historically, these stem from early formulations of strong-coupling perturbation theory,\cite{Keiter_Perturbation_1970,Keiter_Diagrammatic_1971}
    and pioneering works expressed approximations of this type in terms of slave--bosons and later effective pseudoparticles representing many-body states and projected into a physical subspace.\cite{barnes_new_1976,Coleman_New_1984,kuramoto_self_1983,Grewe_Diagrammatic_1981,Grewe_Perturbation_1983,maekawa_crystal_1985}
    These ideas rapidly evolved into self-consistent large-$N$ theories.\cite{bickers_self-consistent_1987,Bickers_Review_1987}
    Finite-$U$ generalizations and conserving extensions broadened the formalism beyond the original setting, which addressed infinitely strong on-site interactions, and clarified the role of crossing diagrams and vertex renormalization.\cite{Pruschke_Anderson_1989,Keiter_The_1990}
    
    As applications within DMFT emerged,\cite{Pruschke_Hubbard_1993} it became increasingly important to understand the limitations of the methods and improve accuracy by including higher-order self-energies and vertex corrections.\cite{Haule_Anderson_2001,kirchner_dynamical_2004,Grewe_Conserving_2008}
    For example, several works closely examined the validity of the Fermi liquid picture at low temperatures, mostly in the single-orbital case\cite{kuramoto_self_1983,muller-hartmann_self-consistent_1984,kroha_unified_1997,kirchner_self-consistent_2002,Kroha_Conserving_2005} while later work focused on high-frequency coefficients of the self-energy and thermodynamic expectation values in two-orbital models.\cite{ruegg_sum_2013}
    
    Neither the inclusion of an additional expansion order (often called the one-crossing approximation, OCA) nor the incorporation of vertex corrections provides a fully satisfactory and systematic picture by itself.\cite{Haule_Anderson_2001, Grewe_Conserving_2008,schmitt_dynamic_2009}
    Including higher orders --- i.e. going to third\cite{Eckstein_Nonequilibrium_2010} or, stochastically, fifth\cite{haule_dynamical_2010} order of some variation of the self-consistent expansion --- improves the behavior at progressively lower temperatures, though at rapidly growing expense.
    Recent work has used decomposition techniques like the discrete Lehmann representation (DLR),\cite{kaye_discrete_2022} tensor train integration \cite{nunez_fernandez_learning_2022,nunez_fernandez_learning_2025} or sum-of-exponential (SOE) representations to significantly accelerate such summations,\cite{kaye_decomposing_2024,yu_inchworm_2025} obtaining accurate fifth- and sixth-order results at massively reduced costs.\cite{yu_inchworm_2025,huang_automated_2025}
    
    In addition to renormalized expansions at low perturbation orders, the underlying hybridization expansions have been employed in very fruitful ways in their bare form within equilibrium applications where the imaginary time Matsubara formalism can be used.
    This allows summing the bare imaginary time expansion to extremely high orders within continuous-time quantum Monte Carlo methods
    \cite{werner_continuous-time_2006,gull_continuous-time_2011} and more recently tensor train techniques.\cite{erpenbeck_tensor_2023}
    Where this approach is not limited by sign problems, it can provide quantitative accuracy.

    On the other hand, diagrammatic Monte Carlo methods are generally substantially more difficult to use in real time, due to dynamical sign problems.\cite{muhlbacher_real-time_2008,werner_diagrammatic_2009,schiro_real-time_2009,schiro_real-time_2010,swenson_application_2011,cohen_memory_2011,swenson_semiclassical_2012,Cohen_Numerically_2013,vanhoecke_diagrammatic_2024}
    Here, several modern pathways forward have emerged from hybridization expansions.
    For example, Inchworm expansions use short-time bare simulations to obtain numerically exact propagators, then extend them to later time by constructing renormalized perturbation expansions that leverage shorter-time propagators to obtain a mitigated sign problem.\cite{Cohen_Taming_2015, Antipov_Currents_2017,Dong_Quantum_2017,Krivenko_Dynamics_2019,Kleinhenz_Dynamic_2020,Atanasova_Correlated_2020,kleinhenz_kondo_2022,Erpenbeck_Quantum_2023,Erpenbeck_Shaping_2023,atanasova_stark_2024,Erpenbeck_Steady_2024,Kunzel_Numerically_2024}
    This can also be effective in imaginary-time expansions where sign problems result from multiorbital structure.\cite{Eidelstein_Multiorbital_2020,goldberger_dynamical_2024,strand_inchworm_2024}
    Within Inchworm methods, fast summation techniques can allow access to very high expansion orders,\cite{rossi_determinant_2017,Boag_Inclusion_2018,yang_inclusion-exclusion_2021}
    and vertex corrections can be incorporated.\cite{kim_pseudoparticle_2022,kim_vertex-based_2023}
    Another set of promising pathways is once again based on the decomposition of high-dimensional integrals appearing within diagrammatic expansions into tractable objects that can be efficiently evaluated without the need for Monte Carlo integration.
    Examples include tensor train techniques\cite{geng_third-order_2025,kim_strong_2025} and SOEs.\cite{paprotzki_high_2025}

    It is interesting at this point to make a connection with a broader literature surrounding quantum impurity models.
    Here, Jianshu Cao, to whom this special issue is dedicated, has made many major contributions.
    Particularly in the context of the spin--boson model,\cite{leggett_dynamics_1987} Cao's group has worked on understanding and extending the regime of validity of low-order perturbative approaches.
    Their work clarified the structure and limitations of second-order quantum master equations --- which can be viewed as a renormalized hybridization expansion in Liouville space --- emphasizing how the choice of representation and system--bath partitioning can qualitatively affect their accuracy.\cite{cao_phase-space_1997-1, cao_effects_2000}
    Building on this perspective, subsequent studies developed and benchmarked reorganized perturbative schemes, most notably polaron and variational polaron transformations, which interpolate between weak- and strong-coupling limits by partially resumming system--bath interactions before truncation.\cite{lee_accuracy_2012,hsieh_nonequilibrium_2019}
    These approaches demonstrated that the apparent breakdown of low-order perturbation theory is often not simply a matter of missing higher-order diagrams, but of working in an inappropriate expansion frame, and that suitable transformations can substantially extend the quantitative reliability of second-order treatments; and have been useful in the NCA context for impurity models mixing fermionic and bosonic degrees of freedom.\cite{chen_anderson-holstein_2016}
    Cao's group also contributed to higher-order studies of the spin--boson model by extension of the Born--Markov expansion to higher orders.\cite{hsieh_unified_2018}
    We note that NCA-style methods and their high-order extensions were applied much later by others to many of the topics explored by the Cao group, including relaxation dynamics in the spin--boson model,\cite{chen_inchworm_2017,chen_inchworm_2017-1,yang_inclusion-exclusion_2021,cai_fast_2022,cai_inchworm_2023,cai_numerical_2023,wang_real-time_2023,wang_solving_2025,goulko_transient_2025,goulko_transient_2026} full counting statistics\cite{Ridley_Numerically_2018,ridley_current_2015,Erpenbeck_Revealing_2020,pollock_reduced_2022,Zemach_Nonequilibrium_2024} and heat transport.\cite{Ridley_Numerically_2019}

    Despite this long history, the behavior of low-order hybridization expansion methods in genuine multi-orbital settings --- and in particular the conditions under which single-orbital intuition carries over --- has received comparatively little systematic attention.
    Here, we revisit the application of low-order NCA-type approximations to multi-orbital systems.
    Although earlier work on multi-orbital NCA-type methods has largely focused on equilibrium and imaginary-time  properties,\cite{ruegg_sum_2013} we address real-frequency spectral functions and nonequilibrium steady-state characteristics.
    Our main finding is that, within low-order hybridization expansions such as NCA and OCA, the ability to capture correlation effects in  multi-orbital systems is governed by the least correlated orbital, as it suppresses the  correlated features of all other orbitals through an unphysical  coupling induced by the truncated expansion.
    We derive this result analytically in the decoupled orbital limit  and confirm it numerically across a range of model parameters, identifying the conditions under which low-order methods of this type can succeed or fail in multi-orbital settings.
    The remainder of this paper is organized as follows. 
    Sec.~\ref{sec:Model} introduces the multi-orbital Anderson impurity model and the decoupled orbital limit that serves as our controlled reference point. 
    Sec.~\ref{sec:Methods} describes the hybridization expansion framework and its NCA and OCA realizations. 
    The results are presented in Sec.~\ref{sec:results}: 
    Sec.~\ref{sec:MathResults} develops the analytic arguments for the behavior of low-order expansions in multi-orbital systems, and Sec.~\ref{sec:NumericalResults} presents the supporting numerical results. 
    Conclusions are given in Sec.~\ref{sec:conclusion}.

\section{Model}\label{sec:Model}

    In this section, we introduce the quantum impurity models considered in this work. We first present the general multi-orbital Anderson impurity model in Sec.~\ref{sec:anderson_model}, and then discuss the separable limit in Sec.~\ref{sec:decoupled_model}, which allows for a systematic analysis and comparison of low-order hybridization expansion approaches for single- and multi-orbital settings.
    
    \subsection{General multi-orbital Anderson impurity model}\label{sec:anderson_model}
    
        We consider a general fermionic quantum impurity model described by the Hamiltonian
        \begin{align}
            H &= H_{\mathrm S} + H_{\mathrm B} + H_{\mathrm{SB}} .
        \end{align}
        Here, $H_{\mathrm S}$ denotes the Hamiltonian of the impurity system, $H_{\mathrm B}$ describes a noninteracting fermionic bath, and $H_{\mathrm{SB}}$ accounts for the coupling between impurity and bath. 
        The most general form with spin-independent two-body interactions is
        \begin{align}
            H_{\mathrm S} &= \sum_{ij\sigma\sigma'} \epsilon_{ij}^{\sigma\sigma'} d_{i\sigma}^\dagger d_{j\sigma'}
            + \frac{1}{2} \sum_{ijkl \atop \sigma\sigma'} U_{ijkl}^{\sigma\sigma'}
            d_{i\sigma}^\dagger d_{k\sigma'}^\dagger d_{l\sigma'} d_{j\sigma} .
            \label{eq:H_S}
        \end{align}
        The operators $d_{i\sigma}^{(\dagger)}$ annihilate (create) an electron with spin $\sigma \in \{\uparrow,\downarrow\}$ in impurity orbital $i$, $\epsilon_{ij}^{\sigma\sigma'}$ denotes the single-particle matrix elements. The tensor $U_{ijkl}^{\sigma\sigma'}$ enumerates two-body matrix elements corresponding to electron--electron interactions, allowing for arbitrary intra- and inter-orbital terms.
        Throughout this work, we set $\hbar = k_{\mathrm B} = e = 1$.
        
        The environment of the impurity, referred to as bath, is modeled as an infinite set of effective noninteracting fermionic orbitals,
        \begin{align}
            H_{\mathrm B} &= \sum_l \sum_{k \in l, \sigma} \epsilon_k^\sigma
            c_{k\sigma}^\dagger c_{k\sigma} ,
            \label{eq:H_B}
        \end{align}
        where the bath degrees of freedom are partitioned into subsets $l$, each of which may be characterized by independent thermodynamic parameters such as temperature or chemical potential. The index $k$ labels the single-particle states within bath $l$, with creation and annihilation operators $c_{k\sigma}^{(\dagger)}$ and energies $\epsilon_k^\sigma$. The coupling between impurity and bath is taken to be bilinear and diagonal in spin,
        \begin{align}
            H_{\mathrm{SB}} &= \sum_l \sum_{k \in l, \sigma, i}
            \left( t_{ki\sigma} c_{k\sigma}^\dagger d_{i\sigma}
            + \mathrm{h.c.} \right),
            \label{eq:H_SB}
        \end{align}
        where $t_{ki\sigma}$ denotes the tunneling amplitude between impurity orbital $i$ and bath state $k$ in bath $l$. Given a system-bath coupling of this form, the influence of the bath on the impurity is fully characterized by the hybridization functions
        \begin{align}
            \Delta_{ij\sigma}^<(t) &= \sum_l \int d\omega\,
            e^{i\omega t}\,
            \Gamma_{ij}^{l\sigma}(\omega)\,
            f_{l\sigma}(\omega),  \\
            \Delta_{ij\sigma}^>(t) &= \sum_l \int d\omega\,
            e^{-i\omega t}\,
            \Gamma_{ij}^{l\sigma}(\omega)\,
            \left[ 1 - f_{l\sigma}(\omega) \right] ,
            \label{eq:def_Delta}
        \end{align}
        with bath occupation functions $f_{l\sigma}(\omega)$ and coupling strength functions
        $\Gamma_{ij}^{l\sigma}(\omega)
            = 2\pi \sum_{k \in l}
            t_{ki\sigma} t_{kj\sigma}^*\,
            \delta(\omega - \epsilon_{k\sigma})$. 
        The hybridization function provides a compact description of the bath.
        It can be used to describe leads,
        to define effective environments in embedding schemes such as dynamical mean field theory,
        \cite{Georges_Dynamical_1996, Freericks_Nonequilibrium_2006, Aoki_Nonequilibrium_2014}
        or to encode microscopic information from a surrounding system.
        \cite{Erpenbeck_Shaping_2023}
        Also, compact representations of the hybridization function or the coupling strength function have proven highly effective in a wide range of applications, enabling both efficient numerical treatments and controlled analytical approximations.\cite{Georges_Dynamical_1996, Arrigoni_Nonequilibrium_2013, Dorda_Auxiliary_2014, Hou_Hierarchical_2014, Cui_Highly_2019, Mejuto_Efficient_2020, Chen_Universal_2022, Baran_Surrogate_2023, Takahashi_High_2024, Florez_Bath_2025, Yu_Multi_2025, Zhang_Minimal_2025}
        Nonequilibrium situations can be imposed naturally by coupling the impurity to baths with different occupation functions $f_{l\sigma}$. For example, a bias voltage arises from baths with different chemical potentials, thermal gradients from baths held at different temperatures, and more general driven environments can be modeled by appropriately chosen nonequilibrium distributions.
        \cite{Kunzel_Numerically_2024}

    \subsection{Decoupled orbital limit and decomposition into multiple independent orbitals}
    \label{sec:decoupled_model}
        In the absence of interactions and hybridization between different impurity orbitals, the model simplifies considerably.
        For notational simplicity and without loss of generality, we assume that the single-particle part of $H_{\mathrm S}$ is diagonal in both orbital and spin indices so that
        $
            \epsilon_{ij}^{\sigma\sigma'}=\epsilon_i^\sigma\delta_{ij}\delta_{\sigma\sigma'}.
        $
        For the impurity to be separable into independent orbitals, we consider the limit in which the Coulomb interaction tensor is diagonal in the orbital indices,
        \begin{align}
            U_{ijkl}^{\sigma\sigma'} = U_i^{\sigma\sigma'}\,
            \delta_{ij}\delta_{ik}\delta_{il},
        \end{align}
        and the hybridization function is orbital diagonal,
        \begin{align}
            \Delta_{ij\sigma}(t) = \delta_{ij}\, \Delta_{i\sigma}(t) .
        \end{align}
        In this case, the impurity Hamiltonian, the bath Hamiltonian, and the system-bath coupling all decompose into independent orbital sectors.
        As a consequence, the full Hamiltonian can be written as a sum of independent single-orbital Anderson impurity models,
        \begin{align}
            H = \sum_i H_i ,
        \end{align}
        where each $H_i$ describes a single interacting impurity orbital coupled to its own effective bath,
        with
        \begin{align}
            H_i &= H_{\mathrm S}^{(i)} + H_{\mathrm B}^{(i)} + H_{\mathrm{SB}}^{(i)} , \\
            H_{\mathrm S}^{(i)} &= \sum_{\sigma} \epsilon_{i\sigma}
            d_{i\sigma}^\dagger d_{i\sigma}
            + U_i
            d_{i\uparrow}^\dagger d_{i\uparrow}
            d_{i\downarrow}^\dagger d_{i\downarrow} .
            \label{eq:Hi_S}
        \end{align}
        Here, $U_i$ denotes the local Coulomb interaction on orbital $i$
        and influence of the bath on the orbital $i$ is fully characterized by the corresponding hybridization function
        $\Delta_{i\sigma}^{\lessgtr}(t)$.
        Since the Hamiltonian is a direct sum of contributions of independent orbtials, the total Hilbert space factorizes into a tensor product of single-orbital subspaces; and as a consequence, any fermionic many-body eigenstate $\ket{\alpha}$ of the full system can be written as a product state in orbital space,
        \begin{align}
        \ket{\alpha} = \bigotimes_i \ket{\alpha_i} ,
        \label{eq:alpha_factorization}
        \end{align}
        where $\ket{\alpha_i} \in \mathcal H_i$ denotes the many body state of orbital $i$ and its associated baths. The composite index $\alpha$ thus represents the collection of orbital resolved quantum numbers, $\alpha = \{\alpha_1, \alpha_2, \ldots\}$.
        
        In this limit, as the many-body problem factorizes into independent single-orbital problems, the entire system can either be described  by applying single-orbital techniques to each orbital individually, or by using a direct multi-orbital formulation.
        Therefore, this decomposition --- which can also be generalized further by including orbital basis changes --- provides a controlled reference point for assessing the role of inter-orbital interactions and hybridizations.
        This can be used to benchmark numerically exact techniques for multi-orbital systems against established single-orbital techniques.\cite{Erpenbeck_Steady_2024}
        Within the scope of this work, we use multi-orbital systems that decouple into independent single-orbital components as a natural benchmark for evaluating the validity of low order hybridization expansion approaches in multi-orbital systems through direct comparison with their well established single-orbital counterparts.

\section{Methods}\label{sec:Methods}

    In this section, we describe the theoretical framework and numerical approach that we use to solve the quantum impurity problem. We first define the nonequilibrium Green's functions and the spectral function in Sec.~\ref{sec:Observables}, which serve as our primary observables and allow us to assess the emergence of sharp Kondo resonances as hallmarks of strongly correlated physics.
    Then, in Sec.~\ref{sec:hybridization_expansion}, we introduce the hybridization expansion and describe its low-order self-consistent realizations: the non-crossing approximation (NCA) and the one-crossing approximation (OCA).
    Finally, we discuss the steady-state limit in Sec.~\ref{sec:steady-state}, which provides the basis for all numerical results presented in Sec.~\ref{sec:results}.
    
    \subsection{Observables of Interest -- Green's Functions and the Spectral Function}\label{sec:Observables}

        The central observable of interest in this work is the energy-resolved single-particle spectral function, which provides direct access to correlation-induced features such as the Kondo resonance.
        In correlated impurity systems, the emergence of narrow peaks often serves as an indicator of strong many-body correlations.
        Predicting the parameters where this type of physics appears can be used to benchmark the accuracy of an impurity solver.
        
        We formulate our theory in real time. 
        The dynamics begins from an initial reference time $t_0$ at which the system is prepared in a many-body state described by the density matrix $\rho(t_0)=\rho_\text{S}\otimes\rho_\text{B}$.
        This initial condition is factorized into a bath part $\rho_\text{B}$ and an impurity part $\rho_\text{S}=\ket{\alpha}\bra{\alpha}$, the latter of which is determined by the impurity state $\ket{\alpha}$. 
        Expectation values of single-time observables are defined as
        $\langle \mathcal{O}(t) \rangle 
        = \mathrm{Tr}\!\left[ \rho(t_0)\, \mathcal{O}(t) \right]$,
        where operators evolve in the Heisenberg picture under the full Hamiltonian for $t \ge t_0$.
        Extending this to two-time correlation functions, the lesser and greater Green's functions are defined as
        \begin{align}
        G^{<}_{i\sigma,j\sigma'}(t,t') 
        &= i \left\langle 
        d^\dagger_{j\sigma'}(t') d_{i\sigma}(t) 
        \right\rangle, \\
        G^{>}_{i\sigma,j\sigma'}(t,t') 
        &= -i \left\langle 
        d_{i\sigma}(t) d^\dagger_{j\sigma'}(t') 
        \right\rangle.
        \end{align}
        These Green's function components encode information about occupations, correlations, and particle exchange processes.
        The retarded Green's function is then defined in terms of the lesser and greater functions:
        \begin{align}
        G^{r}_{i\sigma,j\sigma'}(t,t') 
        = \theta(t-t') 
        \left[
        G^{>}_{i\sigma,j\sigma'}(t,t') 
        - G^{<}_{i\sigma,j\sigma'}(t,t')
        \right],
        \end{align}
        and is directly related to the spectral function (cf.~Eq.~(\ref{eq:def_spec_func})).

        To calculate the real-time Green's functions numerically, we first define the restricted propagators that explicitly trace out the bath degrees of freedom. 
        The single-branch restricted propagator is defined as 
        \begin{align}
        \varphi_\alpha^\beta(t) 
        = \mathrm{Tr}_\mathrm{B} \left\{
        \rho_\mathrm{B}\,
        \langle \beta | e^{-iH(t-t_0)} | \alpha \rangle
        \right\},
        \label{eq:propagator_single}
        \end{align}
        where $\bra{\alpha}$ and $\ket{\beta}$ respectively denote many-body states in the conjugate and normal Hilbert space of the impurity; and $\mathrm{Tr}_\mathrm{B}$ denotes the trace over the bath degrees of freedom, and $t_0$ is the initial time. 
        We further introduce the two-branch restricted propagator spanning both branches of the Keldysh contour,
        \begin{align}
        \Phi_\alpha^{\beta'\beta}(t',t) 
        = \mathrm{Tr}_\mathrm{B} \left\{
        \rho_\mathrm{B}\,
        \langle \alpha | e^{iH(t'-t_0)} | \beta' \rangle
        \langle \beta | e^{-iH(t-t_0)} | \alpha \rangle
        \right\}, \nonumber \\
        \label{eq:propagator_double}
        \end{align}
        where $t$ and $t'$ are real times located on different branches of the Keldysh contour and $\ket{\alpha}$ is the impurity state at the reference time $t_0$.
        Once the restricted propagators are known, the Green’s functions can be constructed. In practice, the corresponding expressions depend on the order of the hybridization expansion and become increasingly lengthy at higher orders, where diagrammatic representations are particularly useful. We provide the explicit relations between Green’s functions and restricted propagators at the NCA and OCA levels in Eqs.~(\ref{eq:G<_from_Phi})--(\ref{eq:G>_from_Phi}) and Eqs.~(\ref{eq:G<_from_Phi_OCA})--(\ref{eq:G>_from_Phi_OCA}).
        These relations make explicit that the evaluation of the Green’s functions reduces to the computation of the restricted propagators $\varphi_\alpha^\beta(t)$ and $\Phi_\alpha^{\beta'\beta}(t',t)$, which is discussed in the following sections.

    \subsection{Propagator-Based Formulation of the Hybridization Expansion} \label{sec:hybridization_expansion}

        We solve the nonequilibrium impurity problem using the hybridization expansion, a perturbative expansion in the system--bath coupling $H_{\mathrm{SB}}$. 
        While one may in principle expand physical observables such as Green's functions directly in powers of $H_{\mathrm{SB}}$, modern NCA-like approaches instead reorganize the expansion into self-consistent schemes that resum infinite classes of diagrams.
        This strategy often ensures that certain properties are conserved in the perturbative treatment, and captures some aspects of correlated physics while remaining computationally tractable.
        In this work, we employ a propagator-based formulation of the hybridization expansion, in which the restricted propagators $\varphi_\alpha^\beta(t)$ and $\Phi_\alpha^{\beta'\beta}(t',t)$ introduced in Eqs.~(\ref{eq:propagator_single}) and (\ref{eq:propagator_double}) are treated as the fundamental dynamical objects.
        The perturbative expansion in $H_{\mathrm{SB}}$ then generates a hierarchy of corrections to these propagators, which can be truncated at different levels of approximation which we will discuss in detail in Sec.~\ref{sec:NCA_OCA}.

        In a propagator-based formulation of the hybridization expansion, the restricted single-branch propagator $\varphi_\alpha^\beta(t)$ can be written as the solution of a Dyson equation,
        \begin{align}
        \label{eq:Dyson_G}
        \varphi_\alpha^\beta(t) 	
        &= \varphi_{0\alpha}^{\beta}(t) \\ 
        -& \sum_{\gamma\gamma'} 
        \int_{t_0}^{t} \hspace{-0.1cm} d\tau_1 
        \int_{t_0}^{\tau_1} \hspace{-0.2cm} d\tau_2 \,
        \varphi_{0\gamma}^\beta(t-\tau_1) 
        \Sigma_{\gamma'}^{\gamma}(\tau_1-\tau_2) 
        \varphi_\alpha^{\gamma'}(\tau_2) , 
        \nonumber
        \end{align}
        where $\varphi_{0\alpha}^{\beta}(t)= \mathrm{Tr}_\mathrm{B} \left\lbrace
        \rho_\mathrm{B}\, \braket{\beta| e^{-i H_\mathrm{S} t}|\alpha} \right\rbrace$
        is the zero-order propagator in the absence of system-bath coupling. The single-branch self-energy $\Sigma_\alpha^\beta(t)$ encodes the influence of the bath on the system dynamics, typically depends on the propagator itself, and when using  a perturbative treatment, its approximate form defines the level of the hybridization expansion. The self-consistent solution of Eq.~\eqref{eq:Dyson_G} thus resums an infinite subset of contributions from the hybridization expansion, consisting of repeated and nested processes of the type encoded in the self-energy.
        Similarly, the two-branch restricted propagator
        $\Phi_\alpha^{\beta'\beta}(t',t)$ obeys a Dyson-type equation that couples
        forward and backward branches of the Keldysh contour,
        \begin{align}
        \label{eq:Dyson_Phi}
        \Phi_\alpha^{\beta'\beta}(t',t)
        &=
        \varphi_\alpha^\beta(t)\,
        \big[\varphi_\alpha^{\beta'}(t')\big]^*
        \\ 
        &+
        \sum_{\gamma\gamma'\delta\delta'}
        \int_{t_0}^{t}  \hspace{-0.1cm} d\tau_1
        \int_{t_0}^{t'} \hspace{-0.1cm} d\tau_1'
        \,
        \varphi_\delta^\beta(t-\tau_1)\,
        \times \nonumber \\ 
        &\times
        \big[\varphi_{\delta'}^{\beta'}(t'-\tau_1')\big]^* 
        \xi_{\gamma\gamma'}^{\delta\delta'}(\tau_1'-\tau_1)\,
        \Phi_\alpha^{\gamma'\gamma}(\tau_1',\tau_1).
        \nonumber
        \end{align}
        Here, $\xi_{\gamma\gamma'}^{\delta\delta'}(t)$ is the cross-branch self-energy
        encoding correlations induced by the bath between the two contour branches.
        The first term corresponds to factorized forward and backward propagation in
        the absence of cross-branch interactions.
        As for the single-branch case, the self-consistent solution of Eq.~(\ref{eq:Dyson_Phi}) resums an infinite subset of contributions in the hybridization expansion, capturing repeated and nested processes mediated encoded in $\xi_{\gamma\gamma'}^{\delta\delta'}(t)$.
        Again, when an approximate expression for $\xi_{\gamma\gamma'}^{\delta\delta'}(t)$ is used, its specific form determines the order of the hybridization expansion included in a calculation.

        Once the restricted propagators $\varphi_\alpha^\beta(t)$ and 
        $\Phi_\alpha^{\beta'\beta}(t',t)$ have been obtained from the self-energies 
        $\Sigma_\alpha^\beta(t)$ and $\xi_{\gamma\gamma'}^{\delta\delta'}(t)$, the 
        single-particle Green's functions can be calculated. 
        If the exact restricted propagators are known, the Green's functions follow 
        directly from Eqs.~(\ref{eq:G<_from_Phi}) and (\ref{eq:G>_from_Phi}). 
        If only approximate expressions for the restricted propagators are available, 
        the system--bath coupling must be incorporated into the Green's functions in a 
        manner consistent with the approximation used for the propagators. 
        
        Approximate propagators obtained from a truncated hybridization expansion do not explicitly contain all contributions arising from the system-bath 
        interaction.
        Consequently, when constructing the Green's functions, care must be taken to ensure that all contributions from the system--bath coupling are accounted for exactly once, thereby avoiding both omissions and double counting of terms from different orders of the hybridization expansion.
        We will give explicit expressions at the NCA and OCA level in the following section.

    \subsection{Low Order Hybridization Expansion Formulations -- NCA and OCA}\label{sec:NCA_OCA}

        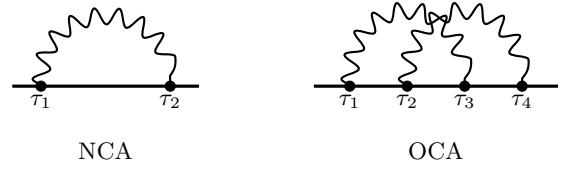
\begin{figure}[tb!]
            \centering
            \begin{tikzpicture}[scale=0.95, baseline=(current bounding box.center)]
            
            \begin{scope}[xshift=-2.3cm]
            
            \draw[very thick] (-1.3,0) -- (1.3,0);
            
            
            \filldraw (-0.9,0) circle (2pt);
            \filldraw (0.9,0) circle (2pt);
            
            \node[below] at (-0.9,0) {$\tau_1$};
            \node[below] at (0.9,0) {$\tau_2$};
            
            \draw[thick, decorate, decoration={snake, amplitude=1.1mm, segment length=3mm}]
            (-0.9,0)
            .. controls (-0.9,1.2) and (0.9,1.2) ..
            (0.9,0);
            
            \node at (0,-0.9) {NCA};
            
            \end{scope}
            
            \begin{scope}[xshift=2.3cm]
            
            \draw[very thick] (-1.7,0) -- (1.7,0);
            
            
            \filldraw (-1.2,0) circle (2pt);
            \filldraw (-0.4,0) circle (2pt);
            \filldraw (0.4,0) circle (2pt);
            \filldraw (1.2,0) circle (2pt);
            
            \node[below] at (-1.2,0) {$\tau_1$};
            \node[below] at (-0.4,0) {$\tau_2$};
            \node[below] at (0.4,0) {$\tau_3$};
            \node[below] at (1.2,0) {$\tau_4$};
            
            \draw[thick, decorate, decoration={snake, amplitude=1.1mm, segment length=3mm}]
            (-1.2,0)
            .. controls (-1.2,1.3) and (0.4,1.3) ..
            (0.4,0);
            
            \draw[thick, decorate, decoration={snake, amplitude=1.1mm, segment length=3mm}]
            (-0.4,0)
            .. controls (-0.4,1.3) and (1.2,1.3) ..
            (1.2,0);
            
            \node at (0,-0.9) {OCA};
            
            \end{scope}
            
            \end{tikzpicture}
            \caption{
            Feynman diagram representation of the contributions retained at the non-crossing approximation (NCA) level (left) and the one crossing approximation (OCA) level (right). Solid lines denote restricted impurity propagators, $\varphi_\alpha^\beta(t)$ or $\Phi_\alpha^{\beta'\beta}(t',t)$, depending on whether one or two branches of the Keldysh contour are involved. Each vertex at time $\tau_i$ corresponds to an action of the system-bath coupling $H_{\mathrm{SB}}$ on the impurity dynamics, while wiggly lines represent hybridization functions encoding the influence of the bath.
            }
            \label{fig:NCA_OCA_diagrams}
        \end{figure}

        After introducing the general framework of the propagator-based hybridization expansion, we provide the explicit expressions for the self-energies and Green's functions at the lowest and second-lowest orders used in this work, namely the NCA and OCA. 
        For the scope of this work, we employ both the NCA and the OCA as representative approaches for low-order self-consistent hybridization expansion methods. While beyond the scope of this work, some of us have benchmarked results from propagator-based NCA and OCA against numerically exact inchworm quantum Monte Carlo calculations,\cite{Erpenbeck_Quantum_2023, Erpenbeck_Steady_2024} providing additional validation of the hybridization expansion in the parameter regimes considered.

        The lowest nonvanishing self-consistent order within the hybridization expansion framework is the non-crossing approximation (NCA).
        \cite{Keiter_Perturbation_1970, Keiter_Diagrammatic_1971, Grewe_Diagrammatic_1981, kuramoto_self_1983, Bickers_Review_1987, Wingreen_Anderson_1994, Haule_Anderson_2001, Eckstein_Nonequilibrium_2010, Gull_Numerically_2011, Roura_Nonequilibrium_2013, Cohen_Greens_2014, Chen_Anderson_2016, Sposetti_Qualitative_2016, Atanasova_Correlated_2020, Erpenbeck_Revealing_2020}
        Within the NCA, all contributions to the restricted propagators can be represented as Feynman diagrams in which hybridization lines do not cross (see Fig.~\ref{fig:NCA_OCA_diagrams}, left). 
        At this level, the self-energies are given by
        \begin{align}
            \Sigma_{\mathrm{NCA}\,\alpha}^{\beta}(t) 
            &= \sum_{ij\sigma \atop \eta = \pm} \sum_{\gamma\gamma'} 
            \Delta_{ij\sigma}^{\eta}(t)
            \braket{\beta|d_{i\sigma}^\eta|\gamma}\,
            \braket{\gamma'|d_{j\sigma}^{\bar\eta}|\alpha}\,
            \varphi_{\gamma'}^{\gamma}(t),
            \nonumber \\
            \label{eq:def_Sigma_G_NCA}
            \\
            \xi_{\mathrm{NCA}\,\gamma\gamma'}^{\delta\delta'}(t)
            &=
            \sum_{ij\sigma} \sum_{\eta=\pm}
            \Delta_{ij\sigma}^{\eta}(t)\,
            \braket{\gamma'|d_{i\sigma}^{\eta}|\delta'}\,
            \braket{\delta|d_{j\sigma}^{\bar\eta}|\gamma}.
            \label{eq:def_xi_NCA}
        \end{align}
        Here, $\eta = +$ corresponds to creation ($d_{i\sigma}^\dagger$) and $\eta = -$ to annihilation ($d_{i\sigma}$) operators, $\bar\eta = -\eta$, and $\Delta_{ij\sigma}^{\mp} = \Delta_{ij\sigma}^{\lessgtr}$. 
        The lesser and greater Green's functions are obtained from the NCA restricted propagators as
        \begin{align}
        G^{<}_{i\sigma,j\sigma'}(t,t') 
        &= i 
        \sum_{\beta\beta'\gamma\gamma'}
        \langle \beta' | d^\dagger_{j\sigma'} | \gamma' \rangle
        \langle \gamma | d_{i\sigma} | \beta \rangle  \times \nonumber \\
        &\times \Phi_\alpha^{\beta'\beta}(t',t)
        \varphi_{\gamma}^{\gamma'}(t'-t), \label{eq:G<_from_Phi} \\
        G^{>}_{i\sigma,j\sigma'}(t,t') 
        &= -i 
        \sum_{\beta\beta'\gamma\gamma'}
        \langle \beta' | d_{i\sigma} | \gamma' \rangle
        \langle \gamma | d^\dagger_{j\sigma'} | \beta \rangle \times \nonumber\\
        &  \times \Phi_\alpha^{\beta'\beta}(t,t')
        \varphi_{\gamma}^{\gamma'}(t-t'), \label{eq:G>_from_Phi}
        \end{align}
        where $\beta, \beta', \gamma$ and $\gamma'$ are again many-body states in the impurity space.
        These expressions imply that any contributions in which hybridization lines cross the creation or annihilation operators appearing in the definition of the Green's function is discarded.
        
        The NCA successfully captures essential aspects of correlated impurity physics, including the formation of local moments and the emergence of a Kondo resonance at temperatures not far below the Kondo scale,\cite{Eckstein_Nonequilibrium_2010, Cohen_Numerically_2013, Cohen_Greens_2014, Sposetti_Qualitative_2016, de_Souza_Melo_Quantitative_2019, Erpenbeck_Revealing_2020} particularly in the large-$U$ limit and at small bias voltages. 
        However, it is well established that NCA does not fully reproduce the correct low-temperature scaling behavior of the Kondo effect. 
        In this regime, vertex corrections associated with crossing hybridization lines become important.
        \cite{Anders_Perturbational_1994, Anders_Beyond_1995, Grewe_Conserving_2008, Eckstein_Nonequilibrium_2010}

        A systematic improvement over NCA is provided by the one crossing approximation (OCA).
        \cite{Pruschke_Anderson_1989, Keiter_The_1990, Pruschke_Hubbard_1993, Haule_Anderson_2001, Gerace_Low_2002, Kroha_Conserving_2005, Eckstein_Photoinduced_2013}
        OCA also incorporates  contributions to the restricted propagators that can be visualized by Feynman diagrams with a single crossing of hybridization lines (see Fig.~\ref{fig:NCA_OCA_diagrams} right), corresponding to the second lowest nonvanishing order in the self-consistent hybridization expansion. 
        At this level, the self-energy is given by
        \begin{align}
        \Sigma_{\mathrm{OCA}\,\alpha}^{\beta}(t) 
        &= \Sigma_{\mathrm{NCA}\,\alpha}^{\beta}(t) \nonumber\\
        & - \sum_{ij\sigma;\, i'j'\sigma'} \sum_{\eta,\eta' = \pm} 
        \sum_{\gamma\gamma'\delta\delta'\nu\nu'} 
        \int_0^t d\tau_3 \int_0^{\tau_3} d\tau_4 \, \nonumber\\
        &\times \Delta_{ij\sigma}^{\eta}(t-\tau_4)\,
        \Delta_{i'j'\sigma'}^{\eta'}(\tau_3) \nonumber\\
        & \times
        \braket{\beta|d_{i\sigma}^\eta|\gamma}
        \braket{\gamma'|d_{i'\sigma'}^{\eta'}|\delta} 
        \braket{\delta'|d_{j\sigma}^{\bar\eta}|\nu} 
        \braket{\nu'|d_{j'\sigma'}^{\bar\eta'}|\alpha} \, \nonumber\\
        &\times\varphi_{\gamma'}^{\gamma}(t-\tau_3)\,
        \varphi_{\delta'}^{\delta}(\tau_3-\tau_4)\,
        \varphi_{\nu'}^{\nu}(\tau_4),
        \label{eq:def_Sigma_G_OCA}
        \end{align}
        where the indices $\gamma,\delta,\nu,\ldots$ run over the many-body states of the system subspace. A corresponding diagrammatic representation is shown in Fig.~\ref{fig:NCA_OCA_diagrams} (right).
        Similar to the single-branch self-energy, the two-branch self-energy 
        $\xi_{\gamma\gamma'}^{\mathrm{OCA}\,\delta\delta'}(t)$ 
        also acquires additional contributions corresponding to diagrams in which the hybridization lines cross
        once. In contrast to the single-branch case, hybridization events may occur on either branch of the contour, leading to multiple distinct time orderings and a substantially increased combinatorial complexity. 
        We therefore refrain from explicit giving the corresponding expression here.
        Likewise, at the OCA level, the Green's functions acquire additional contributions corresponding to diagrams in which one hybridization line crosses the operators in the Green's function. These contributions can be expressed in terms of convolutions of the two-branch and single-branch restricted propagators with the hybridization function,  
        \begin{widetext}
        \begin{small}
        \begin{align}
        \label{eq:G<_from_Phi_OCA}
        &G^{<}_{i\sigma,j\sigma'}(t,t') 
        = i 
        \sum_{\beta\beta'\gamma\gamma'}
        \langle \beta' | d^\dagger_{j\sigma'} | \gamma' \rangle
        \langle \gamma | d_{i\sigma} | \beta \rangle\,
        \Bigg( 
        \Phi_\alpha^{\beta'\beta}(t',t)
        \varphi_{\gamma}^{\gamma'}(t'-t) 
        +
        \sum_{i'j'\sigma'' \atop \eta = \pm} \sum_{\delta\delta'\nu\nu'} 
        \Delta_{i'j'\sigma''}^{\eta}(\tau_1-\tau_2)
        \braket{\delta'|d_{i'\sigma''}^\eta|\delta}\,
        \braket{\nu'|d_{j'\sigma''}^{\bar\eta}|\nu}\,
        \\&
        \times 
        \left(
        \int_{t}^{t'} d\tau_1
        \int_{t_0}^{t} d\tau_2
        \Phi_\alpha^{\beta'\nu}(t',\tau_2)
        \varphi_{\delta'}^{\gamma'}(t'-\tau_1) 
        \varphi_{\gamma}^{\delta}(\tau_1-t) 
        \varphi_{\nu'}^{\beta}(t-\tau_2) 
        - 
        \int_{t'}^{t_0} d\tau_1
        \int_{t}^{t'} d\tau_2
        \Phi_\alpha^{\delta'\beta}(\tau_1, t)
        \varphi_{\gamma}^{\nu}(\tau_2-t) 
        \varphi_{\nu'}^{\gamma'}(t'-\tau_2) 
        \varphi_{\beta'}^{\delta}(\tau_1-t') 
        \right)
        \Bigg)
        \nonumber
        \\
        \label{eq:G>_from_Phi_OCA}
        &G^{>}_{i\sigma,j\sigma'}(t,t') 
        = -i 
        \sum_{\beta\beta'\gamma\gamma'}
        \langle \beta' | d_{i\sigma} | \gamma' \rangle
        \langle \gamma | d^\dagger_{j\sigma'} | \beta \rangle\,
        \Bigg( 
        \Phi_\alpha^{\beta'\beta}(t,t')
        \varphi_{\gamma}^{\gamma'}(t-t') 
        +
        \sum_{i'j'\sigma'' \atop \eta = \pm} \sum_{\delta\delta'\nu\nu'} 
        \Delta_{i'j'\sigma''}^{\eta}(\tau_1-\tau_2)
        \braket{\delta'|d_{i'\sigma''}^\eta|\delta}\,
        \braket{\nu'|d_{j'\sigma''}^{\bar\eta}|\nu}\,
        \\&
        \times 
        \left(
        \int_{t'}^{t} d\tau_1
        \int_{t_0}^{t'} d\tau_2
        \Phi_\alpha^{\beta'\nu}(t,\tau_2)
        \varphi_{\delta'}^{\gamma'}(t-\tau_1) 
        \varphi_{\gamma}^{\delta}(\tau_1-t') 
        \varphi_{\nu'}^{\beta}(t'-\tau_2) 
        - 
        \int_{t}^{t_0} d\tau_1
        \int_{t'}^{t} d\tau_2
        \Phi_\alpha^{\delta'\beta}(\tau_1, t')
        \varphi_{\gamma}^{\nu}(\tau_2-t') 
        \varphi_{\nu'}^{\gamma'}(t-\tau_2) 
        \varphi_{\beta'}^{\delta}(\tau_1-t) 
        \right)
        \Bigg)
        \nonumber
        \end{align}
        \end{small}
        \end{widetext}
        The OCA significantly improves the description of strong correlation effects and restores important aspects of the correct Kondo scaling behavior. 
        In equilibrium, OCA has been shown to capture the qualitative physics of the Mott transition with reasonable accuracy.\cite{Eckstein_Photoinduced_2013, Kleinhenz_Dynamic_2020} 
        More generally, the quality of the approximation is expected to improve systematically with increasing expansion order, and OCA therefore yields more accurate results than NCA in regimes where vertex corrections play a significant role.

    \subsection{Steady-State Solution of the Quantum Impurity Model} \label{sec:steady-state}

        Up to this point, we have described how to calculate the Green's functions and how they evolve in time. While the real-time formulation naturally captures transient dynamics following the switch-on of interactions or couplings at $t_0$, our primary interest lies in the long-time limit, where the system reaches a steady-state. In this regime, memory of the initial preparation is lost, and experimentally relevant stationary observables emerge.
        
        In the steady-state, time-translational invariance is restored, so that all Green's functions and the two-time restricted propagators $\Phi_\alpha^{\beta'\beta}(t,t')$ depend only on the time difference $\Delta t = t - t'$. Formally, this corresponds to taking the limit $t_0 \rightarrow -\infty$ and dropping any dependence on the initial state $\alpha$. 
        Beyond this functional simplification, it is often numerically advantageous to compute steady-state expectation values directly, without performing full real-time propagation. As demonstrated in our recent publications,\cite{Erpenbeck_Resolving_2021, Erpenbeck_Quantum_2023, Erpenbeck_Shaping_2023, Erpenbeck_Steady_2024, Zemach_Nonequilibrium_2024, Kunzel_Numerically_2024} the steady-state can be obtained efficiently via a self-consistency scheme for the two-time restricted propagator that leverages the physical intuition that once the steady-state is reached, propagating forward in time no longer changes any propagator or Green's function. 
        We omit the numerical details here and refer the reader to the relevant literature.\cite{Erpenbeck_Resolving_2021, Erpenbeck_Quantum_2023, Erpenbeck_Steady_2024}
        All numerical results presented in Sec.~\ref{sec:results} are obtained using this scheme.
        
        Another key advantage of the steady-state formulation is that it allows for a direct Fourier transformation to the frequency domain,
        \begin{align}
        G^{x}_{i\sigma,j\sigma'}(\omega) 
        = \int_{-\infty}^{\infty} d(\Delta t)\,
        e^{i\omega \Delta t}\,
        G^{x}_{i\sigma,j\sigma'}(\Delta t),
        \end{align}
        where $x \in \{ <, >, r \}$.
        The energy-resolved spectral function, which is our main observable of interest, is obtained from the retarded component as
        \begin{align}
        A_{i\sigma,j\sigma'}(\omega)
        = -\frac{1}{\pi}\,\text{Im}\, G^{r}_{i\sigma,j\sigma'}(\omega),
        \label{eq:def_spec_func}
        \end{align}
        providing a direct connection between real-time dynamics, steady-state correlations, and experimentally accessible signatures.

\section{Results}\label{sec:results}

    The central result of this work is that the accuracy of low-order hybridization expansions in multi-orbital systems is controlled by the least correlated orbital in the system, i.e., the orbital with the fastest decaying retarded Green’s function.
    Concretely, any orbital with a rapidly decaying restricted propagator --- whether due to large hybridization, high  temperature or any other choice in parameters --- acts as a bottleneck that suppresses correlation-induced features such as the Kondo resonance across the entire multi-orbital calculation, even in orbitals that are themselves strongly correlated. 
    In the following section, we present our results. 
    We focus on multi-orbital --- but separable --- systems, as introduced in Sec.~\ref{sec:decoupled_model}, in order to connect the insights gained from single-orbital calculations with their multi-orbital generalizations.
    
    We begin in Sec.~\ref{sec:MathResults} with several analytic considerations for the restricted propagators and Green's functions in multi-orbital systems, where we first review properties of the exact solution, and then discuss how these properties are modified within a low-order approximate treatment.
    In Sec.~\ref{sec:NumericalResults}, we present numerical results for two representative model systems, each consisting of two decoupled orbitals.
    These are treated using single-orbital techniques for each orbital and compared with a full multi-orbital treatment.
    We focus on the emergence and height of the Kondo resonance as a figure of merit to assess how well a low-order hybridization expansions capture correlation effects in multi-orbital systems, and how well results from single-orbital systems are translatable to the more general multi-orbital case.

    \subsection{Properties of Restricted Propagators in Multi-Orbital Systems in the Decoupled Orbital Limit}\label{sec:MathResults}

        While Sec.~\ref{sec:Model} introduced the general multi-orbital impurity model and Sec.~\ref{sec:Methods} presented the corresponding hybridization expansion framework, including its low-order self-consistent formulations, we now specialize to the limit of effectively decoupled orbitals introduced in Sec.~\ref{sec:decoupled_model}.
        In this limit, explicit analytic expressions connecting the multi-orbital problem to the single-orbital problem can be derived.
        We analyze how the NCA and OCA solutions in the multi-orbital setting differ from the combination of single-orbital solutions, identify the factors that control the accuracy of these low-order approximations, and assess to what extent single-orbital intuition carries over to multi-orbital systems. 
        These results will also be the basis for interpreting the numerical results in Sec.~\ref{sec:NumericalResults}.

        \subsubsection{General statements for exact results}

            When considering a multi-orbital system that can be decomposed into individual, decoupled orbitals, as introduced in Sec.~\ref{sec:decoupled_model}, the full Hamiltonian can be written as a sum of independent single-orbital problems,
            \( H = \sum_i H_i \),
            where each $H_i$ describes a single impurity orbital. 
            In this limit, all restricted propagators factorize into products over their single-orbital parts, and the corresponding Green's functions are identical to their single-orbital counterparts.
            This follows directly from the fact that the individual Hamiltonians commute in the decoupled limit. 
            At the level of the restricted propagators, this is particularly transparent, since they are expectation values of exponentials of this Hamiltonian, which then factorize into products over the individual single-orbital contributions.
            This explicit factorization will be exploited throughout the following sections, where it allows restricted propagators, Green's functions, and self-energies to be constructed in a transparent and computationally efficient manner in the decoupled orbital limit.
            
            To make this point more explicit, let $\varphi_{\alpha_i}^{(i)\, \beta_i}(t)$ and $\Phi_{\alpha_i}^{(i)\, \beta_i'\beta_i}(t,t')$ denote the restricted propagators for the single-orbital system associated with orbital $i$.
            Given these exact single-orbitals propagators, the exact restricted propagator characterizing the composite multi-orbital system is given by
            \begin{align}
                \varphi_\alpha^\beta(t) 
                &= 
                \prod_{i=1}^N \varphi_{\alpha_i}^{(i)\, \beta_i}(t) , 
                \\
                \Phi_\alpha^{\beta'\beta}(t,t')  
                &= 
                \prod_{i=1}^N  \Phi_{\alpha_i}^{(i)\, \beta_i'\beta_i}(t,t') .
            \end{align}
            Note that the restricted propagators satisfy the bounds 
            $|\varphi_{\alpha_i}^{(i)\,\beta_i}(t)| \leq 1$ 
            with the normalization 
            $\varphi_{\alpha_i}^{(i)\,\beta_i}(0)=1$, 
            and 
            $|\Phi_{\alpha_i}^{(i)\,\beta_i'\beta_i}(t,t')| \leq 1$ 
            with 
            $\sum_\beta \Phi_\alpha^{\beta\beta}(t,t)=1$ and
            $\Phi_{\alpha_i}^{(i)\,\beta_i\beta_i}(t,t)=p_{\beta_i}(t)$, 
            where $p_{\beta_i}(t)$ denotes the population of the single-impurity state $\beta_i$ at time $t$. 
            In general, assuming the system thermalizes, the absolute values of both restricted propagators decay with increasing propagation time $t$ or time difference $|t-t'|$, respectively. 
            As a consequence, the corresponding multi-orbital restricted propagators, being products of single-orbital contributions, decay more rapidly with $t$ and $|t-t'|$ than their single-orbital counterparts.
            At first glance, this faster decay appears advantageous for numerical calculations.
            That would be particularly meaningful in strongly correlated systems where long-time tails are difficult to resolve accurately, 
            an issue that has recently been addressed by some of us using data-science-based approaches.\cite{Erpenbeck_Compact_2026} 
            At the same time, however, this apparent advantage is counteracted by the increased hybridization order required to accurately capture correlation effects in compound systems (cf.~Sec.~\ref{sec:NumericalResults}).
            Moreover, these bounds imply that with increasing number of orbitals, the full propagators become increasingly short-ranged in time.

            For the orbital-diagonal Green's functions within the NCA, the factorization condition assumes the explicit form \begin{align}
            G^{<}_{i\sigma,i\sigma'}(t,t') 
            = 
            G^{(i)\, <}_{i\sigma,i\sigma'}(t,t') 
             \prod_{j\neq i} \sum_{\beta_j}\Phi_{\alpha_j}^{(j)\, \beta_j\beta_j}(t',t)
            \varphi_{\beta_j}^{(j)\, \beta_j}(t'-t), \nonumber \\
            \\
            G^{>}_{i\sigma,i\sigma'}(t,t')
            = 
            G^{(i)\, >}_{i\sigma,i\sigma'}(t,t')
             \prod_{j\neq i} \sum_{\beta_j}\Phi_{\alpha_j}^{(j)\, \beta_j\beta_j}(t,t')
            \varphi_{\beta_j}^{(j)\, \beta_j}(t-t').
            \nonumber \\
            \end{align}
            Since these Green's functions only act on orbital $i$, this implies that
            \begin{align}
                \prod_{j\neq i} \sum_{\beta_j}\Phi_{\alpha_j}^{(j)\, \beta_j\beta_j}(t',t)
                \varphi_{\beta_j}^{(j)\, \beta_j}(t'-t) = 1
                \label{eq:GF_unity_condition}
            \end{align}
            for all $t$ and $t'$. 
            While this statement is applicable if the exact restricted propagators are known, recovering this property perturbatively may require high orders within the hybridization expansion in practice.

        \subsubsection{Implications for Multi-Orbital Results within Low-Order Hybridization Expansions such as NCA and OCA}\label{sec:LowOrderLimits}

            We now turn to the behavior of the Green's functions, the restricted propagators, and the relation between them in single-orbital and multi-orbital systems within low-order hybridization expansion methods. In this truncated framework, the exact factorization properties discussed in the previous section generally no longer hold, since different orders of the hybridization expansion are selectively included or omitted. Nevertheless, useful structural differences can still be established, which we discuss in the following.\\

            \paragraph{Difference between single-orbital and multi-orbital descriptions at the propagator level:}
            We begin by introducing notation for the restricted propagators at different levels of approximation for the single-orbital and multi-orbital framework. 
            For orbital $i$ in the corresponding single-orbital impurity problem, we denote by 
            $\varphi_{0,\alpha_i}^{(i)\,\beta_i}(t)$ and 
            $\Phi_{0,\alpha_i}^{(i)\,\beta_i'\beta_i}(t,t')$ 
            the restricted propagators at zero hybridization order, i.e., in the decoupled limit. The corresponding NCA and OCA propagators are denoted by 
            $\varphi_{\mathrm{NCA},\alpha_i}^{(i)\,\beta_i}(t)$ and 
            $\varphi_{\mathrm{OCA},\alpha_i}^{(i)\,\beta_i}(t)$, and analogously for $\Phi$. 
            For the multi-orbital calculation we use 
            $\varphi_{0,\alpha}^{\beta}(t)$, 
            $\varphi_{\mathrm{NCA},\alpha}^{\beta}(t)$, and $\varphi_{\mathrm{OCA},\alpha}^{\beta}(t)$
            and analogously for $\Phi$.
            At zero hybridization order, the multi-orbital propagator factorizes, 
            \begin{align} 
                \varphi_{0,\alpha}^{\beta}(t) = \prod_i \varphi_{0,\alpha_i}^{(i)\,\beta_i}(t),  \label{eq:zero_ord_product}
            \end{align} 
            a direct consequence of the individual Hamiltonians for different orbitals commuting. 
            This property generally does not hold for any other finite hybridization order.
            
            Empirically, one often observes a systematic reduction in the magnitude of the restricted propagators with increasing order of the hybridization expansion, i.e.,
            $
            |\varphi_0| > |\varphi_{\mathrm{NCA}}| > |\varphi_{\mathrm{OCA}}| > \dots.
            $
            However, since NCA and OCA correspond to distinct self-consistent resummations rather than nested truncations of a positive-definite expansion, such a hierarchy is not guaranteed in general. 
            No universal monotonic ordering between different truncation levels can be established, which limits the possibility of making fully general statements outside specific parameter regimes.
            We therefore focus on identifying differences in the diagrams that are included in the single-orbital and the multi-orbital treatment of systems with decoupled orbitals.

\begin{figure*}[t]
            \centering
            \resizebox{\textwidth}{!}{%
            \begin{tikzpicture}[baseline=(current bounding box.center)]
            
            
            \node (panelA) at (0,0) {
            \begin{tikzpicture}[scale=1]
            
                \draw[very thick] (-1.7,0.9) -- (1.7,0.9);
                \draw[very thick] (-1.7,-0.9) -- (1.7,-0.9);
            
                \filldraw (-1.2,0.9) circle (2pt);
                \filldraw (1.2,0.9) circle (2pt);
                \node[below] at (-1.2,0.9) {$\tau_1$};
                \node[below] at (1.2,0.9) {$\tau_4$};
            
                \filldraw (-0.4,-0.9) circle (2pt);
                \filldraw (0.4,-0.9) circle (2pt);
                \node[below] at (-0.4,-0.9) {$\tau_2$};
                \node[below] at (0.4,-0.9) {$\tau_3$};
            
                \draw[thick, decorate, decoration={snake, amplitude=1.1mm, segment length=3mm}]
                (-1.2,0.9) .. controls (-1.2,1.9) and (1.2,1.9) .. (1.2,0.9);
            
                \draw[thick, decorate, decoration={snake, amplitude=0.7mm, segment length=3mm}]
                (-0.4,-0.9) .. controls (-0.4,-0.2) and (0.4,-0.2) .. (0.4,-0.9);
            
                \node[left] at (-1.9,0.9) {orbital $i$};
                \node[left] at (-1.9,-0.9) {orbital $j$};
            
                \node at (0.0,0.3) {\Large $\times$};
            
            \end{tikzpicture}
            };
            
            \node (panelB) [right=0.5cm of panelA] {
            \begin{tikzpicture}[scale=1]
                \draw[very thick] (-1.7,0.9) -- (1.7,0.9);
                \draw[very thick] (-1.7,-0.9) -- (1.7,-0.9);
            
                \filldraw (-1.2,0.9) circle (2pt);
                \filldraw (0.4,0.9) circle (2pt);
                \node[below] at (-1.2,0.9) {$\tau_1$};
                \node[below] at (0.4,0.9) {$\tau_3$};
            
                \filldraw (-0.4,-0.9) circle (2pt);
                \filldraw (1.2,-0.9) circle (2pt);
                \node[below] at (-0.4,-0.9) {$\tau_2$};
                \node[below] at (1.2,-0.9) {$\tau_4$};
            
                \draw[thick, decorate, decoration={snake, amplitude=1.1mm, segment length=3mm}]
                (-1.2,0.9) .. controls (-1.2,1.9) and (0.4,1.9) .. (0.4,0.9);
            
                \draw[thick, decorate, decoration={snake, amplitude=1.1mm, segment length=3mm}]
                (-0.4,-0.9) .. controls (-0.4,0.1) and (1.2,0.1) .. (1.2,-0.9);
            
                \node[left] at (-1.9,0.9) {orbital $i$};
                \node[left] at (-1.9,-0.9) {orbital $j$};
            
                \node at (-0.3,0.3) {\Large $\times$};
            \end{tikzpicture}
            };
            
            \node (panelC) [right=0.5cm of panelB] {
            \begin{tikzpicture}[scale=1]
            
                \draw[very thick] (-1.7,0.3) -- (1.7,0.3);
                \draw[very thick] (-1.7,-0.3) -- (1.7,-0.3);
            
                \filldraw (-1.2,0.3) circle (2pt);
                \filldraw (1.2,0.3) circle (2pt);
                \filldraw (0.4,-0.3) circle (2pt);
                \filldraw (-0.4,-0.3) circle (2pt);
            
                \node[below] at (-1.2,0.3) {$\tau_1$};
                \node[below] at (0.4,-0.3) {$\tau_3$};
                \node[below] at (1.2,0.3) {$\tau_4$};
                \node[below] at (-0.4,-0.3) {$\tau_2$};
            
                \draw[thick, decorate, decoration={snake, amplitude=1.1mm, segment length=3mm}]
                (-0.4,-0.3) .. controls (-0.4,0.8) and (0.4,0.8) .. (0.4,-0.3);
            
                \draw[thick, decorate, decoration={snake, amplitude=1.1mm, segment length=3mm}]
                (-1.2,0.3) .. controls (-1.2,1.6) and (1.2,1.6) .. (1.2,0.3);
            
                \node[left] at (-1.9,0.3) {orbital $i$};
                \node[left] at (-1.9,-0.3) {orbital $j$};
            
            \end{tikzpicture}
            };
            
            \node (panelD) [right=0.5cm of panelC] {
            \begin{tikzpicture}[scale=1]
                \draw[very thick] (-1.7,0.3) -- (1.7,0.3);
                \draw[very thick] (-1.7,-0.3) -- (1.7,-0.3);
            
                \filldraw (-1.2,0.3) circle (2pt);
                \filldraw (0.4,0.3) circle (2pt);
                \filldraw (1.2,-0.3) circle (2pt);
                \filldraw (-0.4,-0.3) circle (2pt);
            
                \node[below] at (-1.2,0.3) {$\tau_1$};
                \node[below] at (0.4,0.3) {$\tau_3$};
                \node[below] at (1.2,-0.3) {$\tau_4$};
                \node[below] at (-0.4,-0.3) {$\tau_2$};
            
                \draw[thick, decorate, decoration={snake, amplitude=1.1mm, segment length=3mm}]
                (-0.4,-0.3) .. controls (-0.4,1.0) and (1.2,1.0) .. (1.2,-0.3);
            
                \draw[thick, decorate, decoration={snake, amplitude=1.3mm, segment length=3mm}]
                (-1.2,0.3) .. controls (-1.2,1.8) and (0.4,1.8) .. (0.4,0.3);
            
                \node[left] at (-1.9,0.3) {orbital $i$};
                \node[left] at (-1.9,-0.3) {orbital $j$};

            \end{tikzpicture}
            };
            
            \draw[dashed, thick] ($(panelB.east)!0.5!(panelC.west)+(0,2)$) --
                                 ($(panelB.east)!0.5!(panelC.west)+(0,-3)$);

            \node[align=center, font=\normalsize, text width=5cm] 
            at (panelA.south) [yshift=-10mm] 
            {diagram \textbf{is included} in the product of single-orbital NCA descriptions};
            \node[align=center, font=\normalsize, text width=5cm] 
            at (panelB.south) [yshift=-10mm] 
            {diagram \textbf{is included} in the product of single-orbital NCA descriptions};
            \node[align=center, font=\normalsize, text width=5cm] 
            at (panelC.south) [yshift=-10mm] 
            {diagram \textbf{is included} in the multi-orbital NCA description as hybridization lines do not cross};
            \node[align=center, font=\normalsize, text width=5cm] 
            at (panelD.south) [yshift=-10mm] 
            {diagram \textbf{is not included} in the multi-orbital NCA description as hybridization lines cross};

            \node[font=\bfseries\normalsize, align=center] 
            at ($(panelA.north)!0.5!(panelB.north)+(0,8mm)$)
            {single-orbital description};
            
            \node[font=\bfseries\normalsize, align=center] 
            at ($(panelC.north)!0.5!(panelD.north)+(0,8mm)$)
            {multi-orbital description};

            \end{tikzpicture}
            }
            \caption{
            Feynman diagram representation of contributions retained and omitted at the NCA level. 
            Left: factorized product of single-orbital descriptions, which applies in the decoupled-orbital limit considered in this work. 
            Right: corresponding multi-orbital formulation for the same system. 
            Solid lines denote restricted impurity propagators, $\varphi_\alpha^\beta(t)$ or $\Phi_\alpha^{\beta'\beta}(t,t')$, depending on whether one or two branches of the Keldysh contour are involved. Each vertex at time $\tau_i$ corresponds to an insertion of the system-bath coupling $H_{\mathrm{SB}}$ in the hybridization expansion, while wiggly lines represent hybridization functions encoding the coupling to the bath.}
            \label{fig:NCA_diagrams_single_VS_multi}
        \end{figure*}
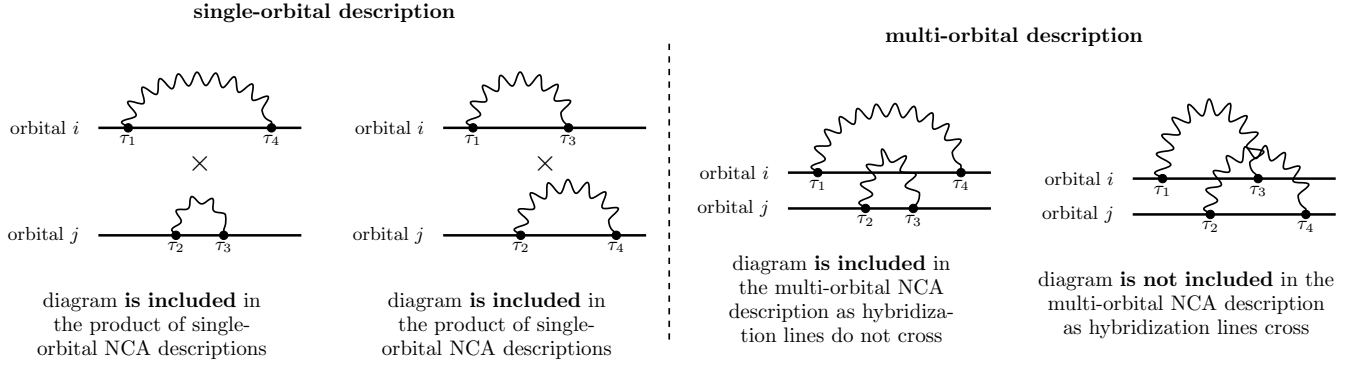

            We now consider a single-orbital NCA treatment that exploits the decoupled orbital structure of our systems.
            The idea is to describe the full system by a product of independent single-orbital propagators. 
            For readability, we define
            \begin{align}
            \zeta_{\mathrm{NCA},\alpha_i}^{(i)\,\beta_i}(t) &= \sum_{\gamma_i\gamma_i'} \int_{t_0}^{t} d\tau_1 \int_{t_0}^{\tau_1} d\tau_2  \\
            &\varphi_{0,\gamma_i}^{(i)\,\beta_i}(t-\tau_1) \Sigma_{\mathrm{NCA}, \gamma_i'}^{(i)\,\gamma_i}(\tau_1-\tau_2) \varphi_{\mathrm{NCA}, \alpha_i}^{(i)\,\gamma_i'}(\tau_2), \nonumber
            \end{align}
            where $\Sigma_{\mathrm{NCA}, \gamma_i'}^{(i)\,\gamma_i}$ denotes the NCA self-energy acting on orbital $i$ only. 
            The corresponding Dyson-like relation for the single-orbital propagator of orbital $i$ then reads
            \begin{align}
                \varphi_{\mathrm{NCA},\alpha_i}^{(i)\,\beta_i}(t) 
                &= 
                \varphi_{0,\alpha_i}^{(i)\,\beta_i}(t) - \zeta_{\mathrm{NCA},\alpha_i}^{(i)\,\beta_i}(t).
            \end{align}
            Within the single-orbital NCA treatment, the restricted propagator of the full system is then given by the product of independent single-orbital NCA contributions,
            \begin{align}
                &\prod_i \varphi_{\mathrm{NCA},\alpha_i}^{(i)\,\beta_i}(t) 
                = 
                \prod_i \left( \varphi_{0,\alpha_i}^{(i)\,\beta_i}(t) - \zeta_{\mathrm{NCA},\alpha_i}^{(i)\,\beta_i}(t) \right)
                \\
                &= \varphi_{0,\alpha}^{\beta}(t)
                - \sum_i
                \left[
                \zeta_{\mathrm{NCA},\alpha_i}^{(i)\,\beta_i}(t)
                \prod_{j\neq i} \varphi_{0,\alpha_j}^{(j)\,\beta_j}(t)
                \right]
                \nonumber \\
                &+ \sum_{i<j}
                \left[
                \zeta_{\mathrm{NCA},\alpha_i}^{(i)\,\beta_i}(t)
                \zeta_{\mathrm{NCA},\alpha_j}^{(j)\,\beta_j}(t)
                \prod_{k\neq i,j} \varphi_{0,\alpha_k}^{(k)\,\beta_k}(t)
                \right]
                - \dots\ ,
                \nonumber
            \end{align}
            where we have used the zero-order factorization property of Eq.~(\ref{eq:zero_ord_product}). 
            This expansion makes explicit that hybridization processes act independently on different orbitals, while terms containing multiple $\zeta_{\mathrm{NCA}}$ factors correspond to combinations of independent hybridization events acting on different orbitals.  
            The corresponding diagrammatic representation is shown in the left panels of Fig.~\ref{fig:NCA_diagrams_single_VS_multi}.
            In a genuine multi-orbital formulation, such contributions would only arise at higher-order hybridization expansions.

            We now turn to the multi-orbital treatment of the same system. In contrast to the single-orbital framework, the multi-orbital NCA propagator is not obtained as a product of orbital-resolved solutions, but rather from a single self-consistent equation defined in the full orbital space. 
            To make this explicit, we introduce
            \(
                \zeta_{\mathrm{NCA},\alpha}^{\beta}(t) 
                = 
                \sum_{\gamma\gamma'} \int_{t_0}^{t} d\tau_1 \int_{t_0}^{\tau_1} d\tau_2 \, \varphi_{0,\gamma}^{\beta}(t-\tau_1) \Sigma_{\mathrm{NCA},\gamma'}^{\gamma}(\tau_1-\tau_2) \varphi_{\mathrm{NCA},\alpha}^{\gamma'}(\tau_2),
            \)
            which can be rewritten into orbital-resolved contributions,
            \(
                \zeta_{\mathrm{NCA},\alpha}^{\beta}(t) = \sum_i \zeta_{\mathrm{NCA},\alpha}^{(i)\,\beta}(t).
            \)
            The restricted propagator of the full system then satisfies the Dyson-like equation
            \begin{align}
                \varphi_{\mathrm{NCA},\alpha}^{\beta}(t) 
                &= 
                \varphi_{0,\alpha}^{\beta}(t) - \sum_i \zeta_{\mathrm{NCA},\alpha}^{(i)\,\beta}(t), \label{eq:Dyson_multi}
            \end{align}
            which makes explicit that the multi-orbital treatment only accounts for one hybridization event at a time for each orbital.
            A key difference from the single-orbital case is that while the multi-orbital formulation can be rewritten in terms of the orbital-resolved contributions, the self-energy for the multi-orbital description depends on the propagator of the full multi-orbital system (see Eq.~(\ref{eq:def_Sigma_G_NCA})), rather than only on the propagator of orbital $i$. 
            Because of this, each hybridization event on any given orbital impacts the propagator across all other orbitals as well. 
            This induces an effective coupling between otherwise decoupled orbital sectors, even in the absence of explicit inter-orbital hybridization in the Hamiltonian. 
            Further, as Eq.~(\ref{eq:Dyson_multi}) only accounts for one hybridization event per orbital at a time, diagrammatic contributions included differ from the single-orbital treatment. 
            While the self-consistent solution of Eq.~(\ref{eq:Dyson_multi}) generates nested sequences that can involve multiple orbitals, hybridization-line crossings between different orbitals are not included at the multi-orbital NCA level. 
            This distinction is illustrated in the right panels of Fig.~\ref{fig:NCA_diagrams_single_VS_multi}. 
            Importantly, there is no nesting relation between the single-orbital and multi-orbital formulation; 
            in particular, the single-orbital NCA result is not recovered from a multi-orbital OCA calculation. 
            Capturing all single-orbital NCA contributions within a multi-orbital framework would, in general, require an expansion to infinite order, or alternatively vertex corrections at an order that grows with the number of orbitals.
            
            The distinct diagrammatic and self-consistent structure of the single- and multi-orbital frameworks make a direct qualitative comparison challenging. 
            In the multi-orbital case, the propagator does not factorize into orbital resolved contributions. 
            Instead, hybridization processes enter through a single self-consistent equation defined in the full orbital space, such that hybridization events on any orbital affect the propagator across all orbitals simultaneously.
            This structure has important consequences for the long time behavior of the restricted propagator and the accessible correlation effects, as one typically expects a slowly decaying oscillatory behavior for strongly correlated systems.
            However, as hybridization events in a multi-orbital treatment act collectively on all orbitals, the decay of the multi-orbital propagator is dominated by the fastest decaying component.
            As a result, the long time behavior of the propagator, and thus the correlation features that can be resolved, are effectively limited by the least correlated orbital in the system.
            
            We note that these considerations extend to cross-contour propagators, with additional bookkeeping due to the inclusion of both single- and cross-branch contributions. 
            However, the diagrammatic differences discussed above and illustrated in Fig.~\ref{fig:NCA_diagrams_single_VS_multi} carry over directly.
            Similar conclusions also apply at the OCA level and beyond. Although the structure becomes more involved as the number of contributing diagrams increases, the product of single-orbital OCA propagators (and, more generally, any finite-order hybridization expansion) still contains a larger set of diagrammatic contributions than the corresponding multi-orbital formulation at the same order, and the least correlated orbital in the system still dominates the correlation features that can be resolved.

            Finally, we reiterate that the need for infinite ladders of corrections to correctly predict certain properties of even the single-orbital problem --- such as the width of the Kondo peak --- was recognized early in the history of NCA-like methods.\cite{Grewe_Conserving_2008}
            Some of the issues in the multi-orbital case can certainly be mitigated by using similar ideas.
            \\

        \paragraph{Difference between single-orbital and multi-orbital descriptions at the Green's function level:}
            When computing Green’s functions from a finite order hybridization expansion, the diagrammatic considerations discussed above carry over. 
            A key point is that the condition Eq.~(\ref{eq:GF_unity_condition}) is only strictly satisfied at infinite order, while  no simple closed expression exists for $\prod_{j\neq i} \sum_{\beta_j}\Phi_{\alpha_j}^{(j),\beta_j}(t',t)\varphi_{\beta_j}^{(j),\beta_j}(t'-t)$ at finite order.
            However, at the NCA level, where the Green’s functions are given by Eqs.~(\ref{eq:G<_from_Phi}) and (\ref{eq:G>_from_Phi}), a simplified expression can be obtained in the limit where all orbitals $j \neq i$ are noninteracting and have zero onsite energy, such that all components of the single- and cross-contour propagators are identical.
            In this case,
            \begin{align}
                \prod_{j\neq i} \sum_{\beta_j}\Phi_{\alpha_j}^{(j),\beta_j\beta_j}(t',t)\varphi_{\beta_j}^{(j),\beta_j}(t'-t) \sim \prod_{j\neq i} G^{(j),<}_{j\sigma,j\sigma}(t,t'),
            \end{align}
            such that
            \begin{align}
                G^{\lessgtr}_{i\sigma,i\sigma'}(t,t')
                &\sim
                G^{(i),\lessgtr}_{i\sigma,i\sigma'}(t,t')
                \times \prod_{j\neq i}
                G^{(j),\lessgtr}_{j\sigma,j\sigma}(t,t').
                \label{eq:GF_factorization}
            \end{align}
            In this limit, the multi-orbital Green’s function factorizes into a product of single-orbital Green’s functions, and Eq.~(\ref{eq:GF_unity_condition}) is satisfied at $t=t'$ only.
            This structure implies again that the decay properties of the full Green’s function are dominated by those of the most rapidly decaying orbital. 
            Since long time tails and oscillatory behavior are again characteristic signatures of strong correlations, such behavior is effectively suppressed by the least correlated orbital within this product structure. 
            Although the applicability of Eq.~(\ref{eq:GF_factorization}) is strongly limited to very specific systems, it provides a reference for interpreting multi-orbital results and highlights how correlation signatures can be damped in finite order hybridization expansions through the presence of rapidly decaying orbitals.

            We note that in higher-order expansions such as the OCA, the diagrammatic structure becomes more complicated, but the same qualitative trends persist.
            Eq.~(\ref{eq:GF_unity_condition}) is satisfied over an extended time range that increases with hybridization order, and achieving full consistency over the relevant $t-t'$ domain generally requires substantially higher expansion orders.
            This is one of the main challenges in multi-orbital calculations, as it directly impacts the ability to resolve correlation-induced features in practical computations.
            For single-orbital product formulations, this limitation does not exist: the explicitly enforced product structure removes the dependence on other orbitals from the Green's function calculation.

    \subsection{Correlation Effects in multi-orbital Systems from Low Order Expansions}\label{sec:NumericalResults}

        This section presents our numerical results, where we investigate under which conditions low order hybridization expansion methods, specifically NCA and OCA, capture correlation effects in multi-orbital impurity systems. As a benchmark for strong correlation physics, we focus on the Kondo resonance. We compare multi-orbital calculations to the exact factorized limit obtained from independent single-orbital impurity problems in order to assess the validity of genuine multi-orbital treatments.
        
        After introducing the model systems in Sec.~\ref{sec:NumericalDetails}, we first study under which conditions correlation effects are captured in two representative setups: systems where a correlated orbital is embedded in a weakly correlated environment, realized here by noninteracting orbitals in Sec.~\ref{sec:result_nonint}, 
        and systems where multiple interacting orbitals generate correlated behavior in Sec.~\ref{sec:result_U1U2}. 
        Across these cases, we systematically vary hybridization strengths to identify regimes in which low order expansions either produce or fail to capture correlated behavior. In Sec.~\ref{sec:result_temp}, we study the effect of temperature on different orbitals and its impact on the resolution of correlation effects. 
        This systematic analysis allows us to identify the parameter regimes of applicability of low order hybridization expansions for multi-orbital systems and to clarify the mechanisms for their success or failure.

        \subsubsection{Model Systems and Numerical Details}\label{sec:NumericalDetails}

            To isolate and study the effects discussed above in a controlled setting, we consider minimal two-orbital impurity models in which the orbitals are strictly decoupled from one another, with each orbital coupled to its own independent bath. In this limit, the exact solution factorizes into a product of two independent single-orbital impurity problems. This factorized solution, evaluated separately for each orbital at the same approximation level (NCA or OCA), serves as a natural benchmark for the corresponding multi-orbital calculation.

            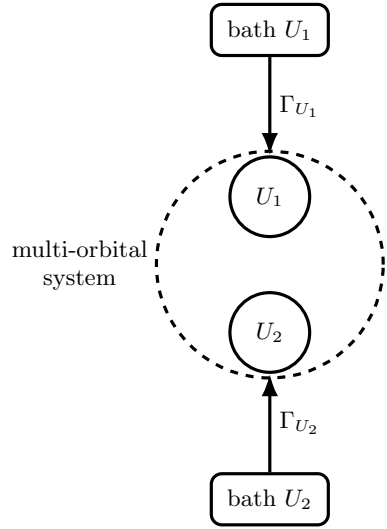
\begin{figure}[t]
            \hspace{-2.25cm}
            \begin{minipage}[t]{0.4\textwidth}
            \begin{tikzpicture}[
                font=\small,
                >=Latex,
                box/.style={draw, rounded corners, very thick, inner sep=6pt, align=center},
                orb/.style={draw, circle, very thick, minimum size=30pt, inner sep=2pt},
                coup/.style={-Latex, very thick},
                note/.style={align=center}
            ]
            
            \node[orb] (int) {$U_1$};
            \node[orb, below=7mm of int] (nonint) {$U_2$};
            
            \coordinate (center) at ($(int)!0.5!(nonint)$);
            
            \draw[very thick, dashed] (center) circle [radius=15mm];
            
            \node[left=15mm of center, align=center]
                {multi-orbital\\system};
            
            \node[box, above=13mm of int] (bathN) {bath $U_1$};
            \node[box, below=13mm of nonint] (bathI) {bath $U_2$};
            
            \draw[coup] (bathN) -- node[right] {$\Gamma_{U_1}$} (int);
            \draw[coup] (bathI) -- node[right] {$\Gamma_{U_2}$} (nonint);
            
            \end{tikzpicture}
            \end{minipage}
            
            \caption{
            Schematic illustration of the model systems used in our numerical calculations, consisting of two spinful orbitals, each coupled to its own bath. 
            The orbital with interaction strength $U_1$ is the primary orbital of interest, while the second orbital with interaction strength $U_2$ serves as an additional spectator orbital, rendering the setup a genuine multi-orbital system. 
            In Sec.~\ref{sec:result_nonint}, the second orbital is taken to be noninteracting ($U_2 = 0$), representing a correlated orbital embedded in a weakly correlated or effectively noninteracting environment. In Sec.~\ref{sec:result_U1U2} and \ref{sec:result_temp}, we instead consider $U_2 = U_1$, corresponding to a system in which multiple interacting orbitals jointly determine the overall correlated behavior.
            }
            \label{fig:cartoon_two_orbitals_multiorbital}
            
            \end{figure}

            We consider the representative model systems schematically illustrated in Fig.~\ref{fig:cartoon_two_orbitals_multiorbital}. The system consists of two completely decoupled spinful orbitals characterized by their respective electron--electron interaction strengths $U_1$ and $U_2$. Throughout this work, we restrict ourself to particle--hole symmetric setups with $\epsilon_{i\sigma} = -U_i/2$.
            The orbital with interaction strength $U_1$ is the primary orbital of interest, while the second orbital with interaction strength $U_2$ acts as a spectator orbital that probes the influence of additional orbitals on the outcome of a multi-orbital calculation.
            Both orbitals are coupled to their own bath, and the absence of any inter orbital coupling, either direct or bath mediated, means that the system factorizes into independent impurity problems.
            This allows for two complementary approaches.
            In a single-orbital treatment (SO), each impurity problem is solved independently within a low order hybridization expansion, and the full solution is obtained from the product of the partial solutions. 
            In contrast, a genuine multi-orbital treatment (MO) solves the coupled problem within a single framework, in such a way that it can be generalized to parameters that do not allow decoupling.
            The decoupled limit therefore provides a controlled setting to quantify deviations between SO and MO approaches at low expansion order and to assess to what extent insights from single-orbital calculations carry over to multi-orbital frameworks.

            Each of the two orbitals is coupled to its own respective independent bath, and baths are described by a wide, flat band with smooth cutoffs,
            \(
            \Gamma_i(\epsilon)
            =
            {\Gamma_i} / 
            \left[{\left(1+e^{\eta(\epsilon-\omega_c)}\right)
             \left(1+e^{-\eta(\epsilon+\omega_c)}\right)}\right]
            \).
            Here $\Gamma_i \in \{\Gamma_{U_1},\Gamma_{U_2}\}$.
            We employ units in which $\hbar=e=1$, and we additionally set $\Gamma=\Gamma_{U_1}$ as our unit of energy.
            This anchors the whole system to the coupling strength of the orbital of interest.
            The coupling to the second orbital is treated as a variable parameter and is explicitly specified for each result.
            The remaining bath parameters are chosen as $\eta=10/\Gamma$ and $\omega_c=25\Gamma$.
            The chemical potential of all baths considered in this work is zero.

            Concerning the numerical details
            and as outlined in Sec.~\ref{sec:Methods}, the spectral functions are obtained from the real-time retarded Green's function directly in the steady-state, using the methodology introduced in Refs.~\onlinecite{Erpenbeck_Quantum_2023,Erpenbeck_Steady_2024}. Although the required integrals can in principle be evaluated by direct quadrature or similar methods, all calculations presented here were performed using quantum Monte Carlo sampling, employing the implementation introduced in Ref.~\onlinecite{Erpenbeck_Steady_2024}. We verified convergence of all restricted propagators with respect to the inchworm time step. 
            In addition, the long-time tails that are characteristic of the restricted propagators in strongly correlated systems were treated using ESPRIT-based extrapolation, as described in Ref.~\onlinecite{Erpenbeck_Compact_2026}.
            All energy-space observables are calculated analytically from this ESPRIT-based representation, as discussed in App.~A in Ref.~\onlinecite{Erpenbeck_Compact_2026}, which significantly improves stability and reproducibility when working with finite-time domain noisy Monte Carlo data.

        \subsubsection{Correlation Effects in Systems with Coexisting Weakly and Strongly Correlated States}\label{sec:result_nonint}

            \begin{figure}[tb]
                \raggedright (a)\\
                \centering
                \includegraphics{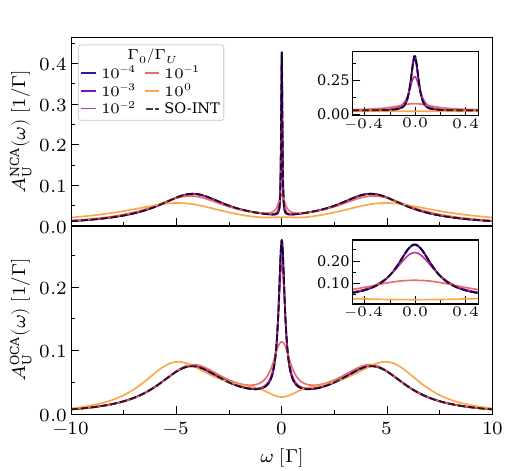}\\
                \raggedright (b)\\
                \includegraphics{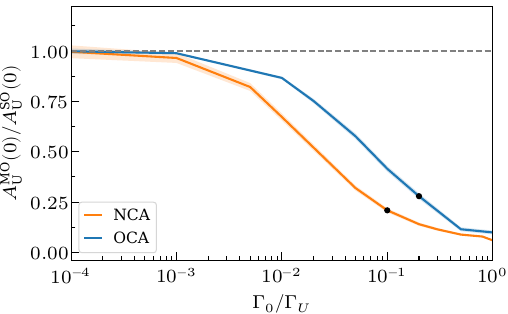}
    
                \caption{
                Spectral function of a two-orbital system consisting of one interacting orbital $U_1 = 8\Gamma$ and one noninteracting orbital $U_2 = 0$, as illustrated in Fig.~\ref{fig:cartoon_two_orbitals_multiorbital}. 
                The hybridization of the noninteracting orbital is varied over $\Gamma_0 = 10^{-4}\Gamma \dots 1.0\Gamma$, while the coupling of the interacting orbital is kept fixed at $\Gamma_U = \Gamma$. The inverse temperature of the system is $\beta = 100/\Gamma$.
                (a) Spectral function of the interacting orbital from a multi-orbital calculation. Top: NCA results. Bottom: OCA results. The dashed black line denotes the corresponding single-orbital calculation.
                Insets provide a close up on the Kondo peak.
                (b) Kondo peak height as a function of \( \Gamma_0 \). Orange and blue curves show NCA and OCA results, and the line thickness indicates the numerical error. 
                The horizontal dashed line indicates the single-orbital value. Black dots mark the onset of a visible peak at \( \Gamma_0 = 0.1\Gamma \) (NCA) and \( 0.2\Gamma \) (OCA).
                } 
                \label{fig:interacting_noninteracting_spectral_nca_oca}
            \end{figure}

            We first study how accurately multi-orbital calculations capture correlation effects in systems consisting of a small number of correlated orbitals coexisting alongside weakly or noninteracting orbitals, as commonly encountered in downfolding and active subspace approaches, where a correlated subspace coexists with weakly correlated or effectively noninteracting orbitals within the same system.
            To this end, we consider the minimal two orbital model shown in Fig.~\ref{fig:cartoon_two_orbitals_multiorbital}, where the first orbital is interacting with $U_1= 8\Gamma$, while the second orbital is noninteracting, $U_2 = 0$.
            Our focus is on the influence of the noninteracting orbital on observables associated with the interacting orbital. Since the orbitals are physically decoupled, any such influence constitutes a spurious effect arising from the low order hybridization expansion. 
            For clarity, we adopt a simplified notation in this section, denoting quantities associated with the interacting orbital by the subscript $U$ and those of the noninteracting orbital by $0$. In particular, $\Gamma_U \equiv \Gamma$ defines the coupling strength of the interacting orbital and sets the unit of energy, while $\Gamma_0$ denotes the coupling strength of the noninteracting orbital, which is variable. The inverse temperature of the entire system is fixed to $\beta = 100/\Gamma$.
            
            To assess how accurately multi-orbital calculations capture correlation effects, we analyze the spectral function of the interacting orbital shown in Fig.~\ref{fig:interacting_noninteracting_spectral_nca_oca}(a). As a key indicator of strong correlations, we focus on the Kondo resonance at $\omega = 0$. The figure displays NCA (top panel) and OCA (bottom panel) results for a wide range of hybridization strengths of the noninteracting orbital, $\Gamma_0$, indicated by the colored curves. The black dashed line corresponds to the single-orbital reference calculation for the interacting orbital at the respective NCA or OCA level and serves as a benchmark for assessing the validity of the multi-orbital treatment.
            
            For all coupling strengths, the spectra exhibit Hubbard sidebands located near $\pm U/2$, reflecting atomic excitations. As the hybridization $\Gamma_0$ of the noninteracting orbital is reduced, a Kondo resonance emerges at low frequencies in both NCA and OCA and becomes progressively sharper as $\Gamma_0$ decreases. At the same time, the spectra approach the single-orbital reference. 
            In the opposite regime, when $\Gamma_0$ becomes comparable to $\Gamma_U$, the Kondo resonance is strongly suppressed and can disappear altogether. For the OCA results, we even observe the emergence of a Kondo-like dip centered at $\omega = 0$, indicating that low order methods can not only fail to capture correlation effects in certain parameter regimes, but may also predict qualitatively incorrect behavior.
            Since the two orbitals are physically decoupled, the observed dependence on $\Gamma_0$, in particular the modification of the Kondo resonance, is entirely artificial and originates from the low order hybridization expansion, which induces an unphysical coupling between the orbitals (see Sec.~\ref{sec:LowOrderLimits}).
            
            To systematically investigate under which conditions NCA and OCA can capture correlation effects in multi-orbital systems, and to determine their parameter regime of applicability, we study the height of the Kondo peak of the interacting orbital as a function of the coupling $\Gamma_0$. 
            The results are shown in Fig.~\ref{fig:interacting_noninteracting_spectral_nca_oca}(b), where the blue and orange lines correspond to multi-orbital calculations, the black markers indicate the onset of a visible Kondo resonance in the spectra, and the grey dashed line indicates the height of the Kondo resonance from  a single-orbital calculation for reference.
            A Kondo resonance consistent with the single-orbital limit is recovered only when $\Gamma_0$ is at least three orders of magnitude smaller than $\Gamma_U$.
            Conversely, for $\Gamma_0$ larger than approximately one order of magnitude below $\Gamma_U$, the Kondo peak is strongly suppressed to the point that it disappears.
            Generally, both NCA and OCA underestimate the Kondo peak height as compared to the single-orbital result.
            OCA consistently improves upon NCA by yielding larger peak heights at fixed coupling and by approaching the benchmark behavior at larger $\Gamma_0$ regime. 
            This systematic improvement suggests a route toward restoring the correct behavior via higher order hybridization expansion methods, albeit at increasing computational cost. Overall, these results support the interpretation that the correct decoupled limit is recovered only upon sufficiently high expansion order.

            \begin{figure}[tb]
                \includegraphics{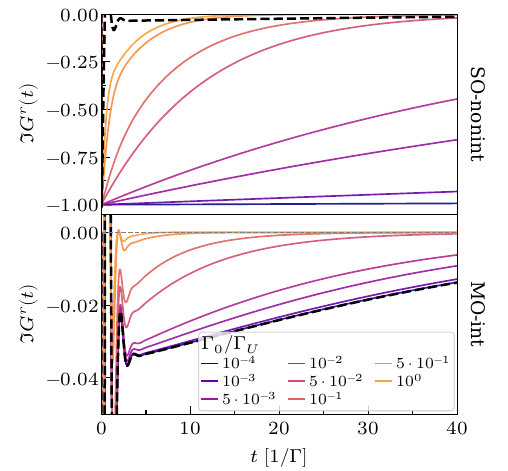}
                \caption{
                Retarded Green’s functions at the NCA level for the same system and parameters as shown in Fig.~\ref{fig:interacting_noninteracting_spectral_nca_oca}. 
                Top: noninteracting orbital for varying \( \Gamma_0 \) from the single-orbital calculation. 
                Bottom: interacting orbital from the multi-orbital calculation for the same parameters.
                In both panels, the dashed black line corresponds to the single-orbital result for the interacting orbital which serves as a reference point. 
                }
                \label{fig:interacting_noninteracting_spectral_nca_oca_GF}
            \end{figure}

            To better understand the dependence of the Kondo peak on $\Gamma_0$, we analyze the imaginary part of the real time retarded Green’s function, shown in Fig.~\ref{fig:interacting_noninteracting_spectral_nca_oca_GF}. 
            The top panel displays the retarded Green’s function of the noninteracting orbital obtained from single-orbital calculations for different values of $\Gamma_0$, 
            providing a reference for the intrinsic effect of hybridization.
            The bottom panel shows the interacting orbital's retarded Green’s function from the multi-orbital calculation, with colors indicating $\Gamma_0$. In both panels, the dashed black line corresponds to the single-orbital result for the interacting orbital which serves as a point of reference.
            For the noninteracting orbital, increasing $\Gamma_0$ leads to a progressively faster decay of the Green’s function. 
            The interacting orbital in the multi-orbital calculation exhibits the same trend: its Green’s function decays more rapidly with increasing $\Gamma_0$ compared to the single-orbital reference. 
            Since the long time decay of the Green’s function determines the spectral weight at $\omega = 0$, this accelerated decay directly results in a suppression of the Kondo resonance.
            This behavior is consistent with Sec.~\ref{sec:LowOrderLimits}, where the correlations captured by low order multi-orbital calculations are determined by the least correlated, i.e., fastest decaying, orbital.
            For large $\Gamma_0$, the noninteracting orbital sets the decay scale, inducing an artificial suppression of the interacting orbital's Green's function and thus reducing the Kondo peak. 
            In contrast, for $\Gamma_0 \to 0$, the noninteracting orbital effectively decouples and its Green’s function approaches unity. In this limit,  the interacting orbital recovers the single-orbital behavior, as it alone determines the relevant timescale.

    \subsubsection{Correlation Effects in Systems Comprising Multiple Correlated States}\label{sec:result_U1U2}

        We now study how accurately multi-orbital calculations capture correlation effects in systems consisting of several correlated orbitals with comparable interaction strengths, as commonly encountered in the description of multiband correlated systems within embedding schemes such as DMFT.
        To this end, we consider the minimal two-orbital model shown in Fig.~\ref{fig:cartoon_two_orbitals_multiorbital}, where both orbitals are interacting with $U_1 = U_2 = 8\Gamma$. 
        As in the previous section, we mainly focus on correlation effects in the first orbital, where $\Gamma_{U_1} \equiv \Gamma$ sets the energy scale, while $\Gamma_{U_2}$ is the coupling strength of the second orbital, which is varied. 
        Since the orbitals are physically decoupled, any dependence of orbital one on $\Gamma_{U_2}$ is a effect purely arising from the low order hybridization expansion methodology. 
        The inverse temperature of the entire system is fixed to $\beta = 100/\Gamma$.

        \begin{figure}[tb]
            \raggedright (a)\\
            \centering
            \includegraphics{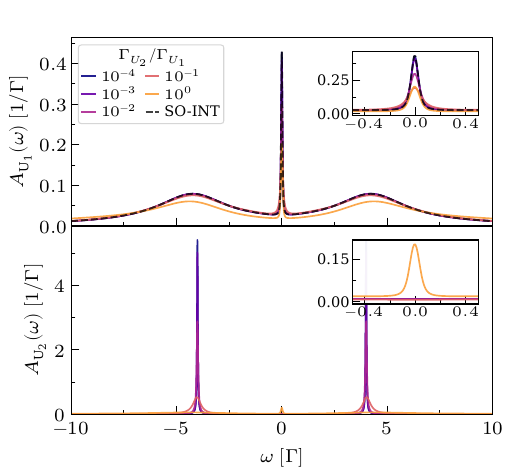}\\
            \raggedright (b)\\
            \includegraphics{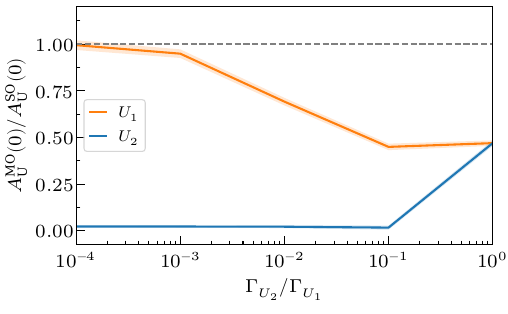}
    
            \caption{
                Spectral function of a two-orbital system consisting of two interacting orbitals with $U_1 = U_2 = 8\Gamma$, as illustrated in Fig.~\ref{fig:cartoon_two_orbitals_multiorbital}, calculated at the NCA level. 
                The hybridization of the the second orbital is varied over $\Gamma_{U_2} = 10^{-4}\Gamma \dots 1.0\Gamma$, while the coupling of the interacting orbital is kept fixed at $\Gamma_{U_1} = \Gamma$. 
                The inverse temperature of the system is $\beta = 100/\Gamma$.
                (a) Spectral function for both orbitals from a multi-orbital calculation. 
                Top: Results for the first orbital $U_1$ with constant coupling to the baths. 
                Bottom: Results for the second orbital $U_2$ with variable coupling to the baths $\Gamma_{U_2}$. 
                Insets provide a close up on the Kondo peak.
                (b) Kondo peak height of both orbtials as a function of the coupling of the second orbital \( \Gamma_{U_2} \). 
                Orange and blue curves show the results for the orbitals $U_1$ and $U_2$, respectively, and the line thickness indicates the numerical error. 
                The horizontal dashed line indicates the single-orbital value.
                }
            \label{fig:spectral_interacting_nca}
        \end{figure}

        To assess how accurately low hybridization order calculations capture correlation effects in multi-orbital systems, we again analyze the Kondo resonance at $\omega = 0$ in the spectral functions calculated at the NCA level as shown in Fig.~\ref{fig:spectral_interacting_nca}(a). 
        The top panel displays the spectral function of the first orbital, while the bottom panel shows that of the second orbital. The colored curves correspond to different coupling strengths $\Gamma_{U_2}$, and the black dashed line in the top panel denotes the single-orbital benchmark for the first orbital.
        Focusing on the first orbital (top panel), we find that the single-orbital benchmark is again recovered in the limit $\Gamma_{U_2} \to 0$. With increasing $\Gamma_{U_2}$, all features in the spectral function broaden and the Kondo peak is reduced, indicating a spurious influence of the second orbital. While this trend is consistent with Sec.~\ref{sec:result_nonint}, the suppression is less pronounced and the peak does not vanish entirely.
        Turning to the second orbital (bottom panel), we find that for small $\Gamma_{U_2}$ the spectral function is dominated by sharply localized peaks associated with atomic excitations. 
        As $\Gamma_{U_2}$ increases, a Kondo peak gradually develops and grows in height (see inset). 
        In the symmetric limit $\Gamma_{U_2} = \Gamma_{U_1}$, the two orbitals exhibit identical spectral features.

        To systematically assess under which conditions the NCA captures correlation effects in multi-orbital systems, we again analyze the height of the Kondo peak as a function of the coupling $\Gamma_{U_2}$. The results are shown in Fig.~\ref{fig:spectral_interacting_nca}(b), where the orange and blue curves denote the spectral weight at $\omega = 0$ for the first and second orbital, respectively, and the grey dashed line indicates the single-orbital reference value.
        We find that the single-orbital benchmark is recovered only when $\Gamma_{U_2}$ is at least three orders of magnitude smaller than $\Gamma_{U_1}$. As $\Gamma_{U_2}$ increases, the Kondo peak is gradually suppressed, although less rapidly than in the case with a noninteracting second orbital discussed in the previous section. In the symmetric limit $\Gamma_{U_2} = \Gamma_{U_1}$, both orbitals behave identically and retain a finite, albeit reduced, Kondo peak.

        \begin{figure}[tb]
            \centering
            \includegraphics{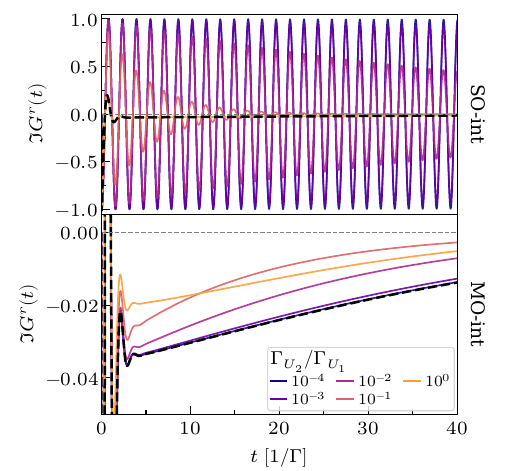}
            \caption{
            Retarded Green’s functions at the NCA level for the same system and parameters as in Fig.~\ref{fig:spectral_interacting_nca}. 
            Top: results for the interacting orbital $U_2$ for varying \( \Gamma_{U_2} \) from the single-orbital calculation. 
            Bottom: results for the interacting orbital $U_1$ from the multi-orbital calculation for the same parameters.
            In both panels, the dashed black line corresponds to single-orbital result for the interacting orbital $U_1$ which serves as a reference point. 
            }
            \label{fig:spectral_interacting_nca_GF}
        \end{figure}

        To better understand this behavior, we analyze the imaginary part of the real-time retarded Green’s function shown calculated at the NCA level, as shown in Fig.~\ref{fig:spectral_interacting_nca_GF}. 
        The top panel displays the Green’s function of the second orbital obtained from a single-orbital calculations for different values of $\Gamma_{U_2}$, serving as a reference for its intrinsic dynamics. 
        The bottom panel shows the Green’s function of the first orbital from the multi-orbital calculation, with colors indicating different values of $\Gamma_{U_2}$. 
        In both panels, the dashed black line corresponds to the single-orbital result for the first orbital, used as a benchmark.
        The Green’s function of the second orbital (top panel) exhibits slowly decaying oscillatory behavior characteristic of interacting dynamics, with the decay of the Green’s function becoming progressively faster as $\Gamma_{U_2}$ increases. As the second orbital is interacting, it exhibits correlated behavior, particularly at larger $\Gamma_{U_2}$, which results in the Green’s function in this regime decaying more slowly than in the noninteracting case discussed in the previous section.
        This slower decay directly affects the multi-orbital calculation, where the overall decay rate is set by the fastest decaying orbital. As shown in the bottom panel, the Green’s function of the first orbital is therefore less strongly suppressed than in the noninteracting case, resulting in a slower reduction of the Kondo resonance with increasing $\Gamma_{U_2}$. While the single-orbital benchmark is still strictly recovered for very small $\Gamma_{U_2}$ only, we note that the interacting nature of the second orbital still allows the multi-orbital NCA calculation to produce a Kondo peak, albeit with an incorrect height.
        These results confirm our interpretation that, within low order hybridization expansions, the ability to capture correlated behavior is governed by the least correlated orbital, i.e., the one with the fastest decaying Green’s function.

    \subsubsection{Temperature Dependence of Correlation Effects in Multi-Orbital Systems}\label{sec:result_temp}

        To further assess the applicability of low order hybridization expansion approaches and to verify our interpretation that the least correlated orbital determines to what extent correlation behavior can be observed in a multi-orbital calculation, we now examine the role of temperature by allowing different temperatures for different orbitals. 
        This extends our analysis beyond the variations in coupling strengths considered in the previous sections, which directly enter the hybridization expansion, to a parameter intrinsically linked to correlation effects.
        As in the previous section, we consider the minimal two-orbital model shown in Fig.~\ref{fig:cartoon_two_orbitals_multiorbital}, with $U_1 = U_2 = 8\Gamma$. 
        As before, we focus on observables of the first orbital, where $\Gamma_{U_1} \equiv \Gamma$ sets the energy scale, and consider two cases for the second orbital: $\Gamma_{U_2} = \Gamma_{U_1}$ and $\Gamma_{U_2} = 0.001\Gamma_{U_1}$, corresponding to equally strong and strongly suppressed hybridization case relative to the first orbital, respectively. The inverse temperature of the first orbital is fixed at $\beta_{U_1} = 100/\Gamma$, while $\beta_{U_2}$ is varied from the low temperature correlated regime to the high temperature uncorrelated regime.
        Since the orbitals are physically decoupled, any dependence of observables in the first orbital on $\beta_{U_2}$ arises from the low order hybridization expansion.

        \begin{figure}[tb]
            \raggedright (a)\\
            \centering
            \includegraphics[width=\columnwidth]{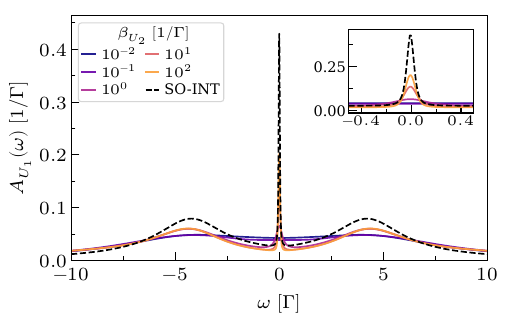}\\
            \raggedright (b)\\
            \centering
            \includegraphics[width=\columnwidth]{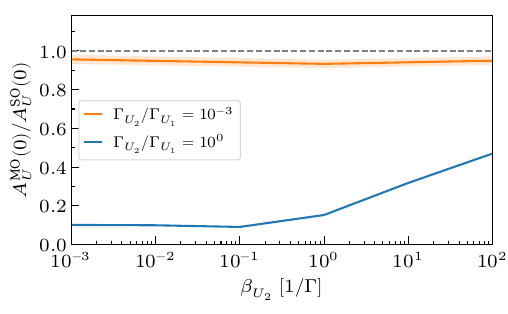}
            \caption{
                Spectral function of a two-orbital system consisting of two interacting orbitals with $U_1 = U_2 = 8\Gamma$, as illustrated in Fig.~\ref{fig:cartoon_two_orbitals_multiorbital}, calculated at the NCA level. 
                The hybridization strengths of both orbitals are fixed. 
                The inverse temperature of the second interacting orbital is varied over the range $\beta_{U_2} = 0.001/\Gamma \dots 100/\Gamma$, while the inverse temperature of the first orbital is kept fixed at $\beta_{U_1} = 100/\Gamma$.
                (a) Spectral function of the first orbital from the multi-orbital calculation for $\Gamma_{U_1} = \Gamma_{U_2}$. Insets show a close up of the Kondo peak, black dashed line is the result from a single-orbital calculation.
                (b) Kondo peak height of the first orbital as a function of the inverse temperature $\beta_{U_2}$. Blue and orange curves correspond to two representative coupling scenarios, one with a strongly coupled second orbital and one with a weakly coupled one. The line thickness indicates the numerical error.}
            \label{fig:spectra_many_beta}
        \end{figure}

        We again assess the capability of low order hybridization expansion methods to capture correlation effects in multi-orbital systems by analyzing the Kondo resonance of the first orbital at $\omega = 0$ in the spectral functions calculated at the NCA level shown in Fig.~\ref{fig:spectra_many_beta}(a). 
        The curves of different color correspond to different temperatures of the second orbital $\beta_{U_2}$, while the grey dashed line denotes the single-orbital benchmark for the first orbital.
        We find that $\beta_{U_2}$ has a pronounced impact on the spectrum of the first orbital and its Kondo peak. At high temperatures of the second orbital, that is small $\beta_{U_2}$ where it remains in the uncorrelated regime, no Kondo peak is visible in the spectrum of the first orbital. As the temperature of the second orbital is decreased, that is with increasing $\beta_{U_2}$ where the second orbital enters the correlated regime, a Kondo peak gradually develops in the first orbital.
        While the single-orbital benchmark is not recovered for the range of $\beta_{U_2}$ considered here, the observed trend indicates that it would be approached only at significantly larger $\beta_{U_2}$. This demonstrates that low order hybridization expansion methods recover correlation effects only when all orbitals are deep in the strongly correlated regime.

        \begin{figure}[tb]
            \centering
            \includegraphics{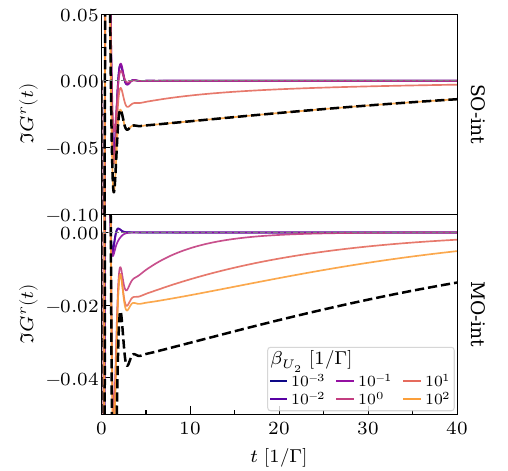}
            \caption{
            Retarded Green’s functions at the NCA level for the same system and parameters as in Fig.~\ref{fig:spectra_many_beta}(a), that is for $\Gamma_{U_1} = \Gamma_{U_2}$. 
            Top: results for the interacting orbital $U_2$ for varying \( \beta_{U_2} \) from the single-orbital calculation. 
            Bottom: results for the interacting orbital $U_1$ from the multi-orbital calculation for the same parameters.
            In both panels, the dashed black line corresponds to single-orbital result for the interacting orbital $U_1$ which serves as a reference point. 
            }
            \label{fig:spectral_interacting_nca_GF_gamma1}
        \end{figure}

        To quantify this temperature dependence, Fig.~\ref{fig:spectra_many_beta}(b) shows the Kondo peak height of the first orbital as a function of $\beta_{U_2}$ for two representative hybridization strengths of the second orbital, corresponding to strongly and weakly coupled cases.
        For the strongly coupled case, $\Gamma_{U_2} = \Gamma_{U_1}$, which corresponds to the data shown in Fig.~\ref{fig:spectra_many_beta}(a), the Kondo peak emerges for $\beta_{U_2} \gtrsim 1.0/\Gamma$. 
        This coincides with the second orbital entering the correlated regime, reflected in a slower decay of its Green’s function (cf.~Fig.~\ref{fig:spectral_interacting_nca_GF_gamma1}).
        In contrast, for the weakly coupled case, $\Gamma_{U_2} = 0.001\Gamma_{U_1}$, the Kondo peak of the first orbital shows no dependence on $\beta_{U_2}$. 
        This is consistent with our interpretation: in this limit, the second orbital is effectively decoupled, and its Green’s function decays on a much longer timescale than that of the first orbital. 
        As a result, varying $\beta_{U_2}$ does not influence the first orbital on the timescales relevant to its dynamics, and its Kondo resonance remains unaffected.

        This point is further supported by analyzing the imaginary part of the real-time retarded Green’s function calculated at the NCA level, shown in Fig.~\ref{fig:spectral_interacting_nca_GF_gamma1} for the strongly coupled case $\Gamma_{U_2} = \Gamma_{U_1}$.
        The top panel shows the Green’s function of the second orbital obtained from single-orbital calculations for different inverse temperatures $\beta_{U_2}$, while the bottom panel shows the Green’s function of the first orbital from the multi-orbital calculation, with colors indicating different values of $\beta_{U_2}$. 
        In both panels, the dashed black line corresponds to the single-orbital reference result for the first orbital.
        The Green’s function of the second orbital (top panel) decays rapidly at high temperature and develops a slowly decaying tail at low temperature, indicative of this orbital entering the correlated regime. 
        This emerging long time tail is reflected in the multi-orbital Green’s function of the first orbital, consistent with the development of the Kondo peak observed in Fig.~\ref{fig:spectra_many_beta}. 
        These results provide further evidence for our interpretation that, within low order hybridization expansion frameworks, the ability to capture correlated behavior is governed by the least correlated orbital, i.e., the orbital with the fastest decaying Green’s function, rather than being determined solely by hybridization strength.

\section{Conclusion}\label{sec:conclusion}

    In this work, we have investigated the accuracy and limitations of low-order hybridization expansion methods, \
    specifically the non-crossing approximation (NCA) and the one-crossing approximation (OCA),
    for multi-orbital quantum impurity models in the 
    nonequilibrium steady-state.
    Using the decoupled orbital limit as a controlled reference point, we derived explicit analytic results connecting multi-orbital restricted propagators and Green's functions to their single-orbital counterparts, and identified the diagrammatic mechanisms responsible for the breakdown of low-order methods in multi-orbital settings.

    Our central finding is that the accuracy of low-order hybridization expansions in multi-orbital systems is determined by the least correlated orbital.
    Any orbital with a rapidly decaying restricted propagator --- whether due to weak effective correlations, strong hybridization, high temperature or other parameter choices than shorten its memory --- is coupled in an unphysical way to all other orbitals through the truncated expansion, suppressing correlation-induced features such as the Kondo resonance even in orbitals that are themselves strongly 
    correlated.
    This effect is not an artifact of a particular parameter choice, but a structural consequence of working at finite expansion order in a multi-orbital framework.

    We demonstrated this mechanism numerically across three representative scenarios.
    First, in systems where a correlated orbital coexists with a 
    noninteracting one, the Kondo resonance of the interacting orbital is strongly suppressed as the hybridization of the noninteracting orbital increases, where qualitatively incorrect behavior can emerge.
    Second, in systems where multiple interacting orbitals are present, we observe the same effect, but the suppression is less severe, since the interacting nature of all orbitals means that their Green's functions decay more slowly. 
    Third, varying the temperature of individual orbitals reveals the  same trend, correlation effects in one orbital are resolved only when all other orbitals are themselves in the correlated regime.
    We further find that OCA systematically improves upon NCA, consistent with the expectation that higher expansion orders provide more accurate results.

    These results have direct implications for practical applications of low-order hybridization expansion methods in multi-orbital settings, such as DMFT impurity solvers for multi-band systems or active subspace approaches.
    In such contexts, the presence of even a single weakly correlated or high-temperature orbital can significantly degrade the accuracy of the calculation for all other orbitals.
    Our analytic results make clear that resolving this issue within a  genuine multi-orbital framework requires expansion orders that grow with the number of orbitals.
    The decoupled orbital limit introduced here provides a natural and controlled benchmark for assessing such improvements, and for validating numerically exact multi-orbital methods against their well-established single-orbital counterparts.
    More broadly, this work contributes to the ongoing effort of  understanding the structural limitations of low-order perturbative  approaches in quantum impurity problems --- a question that has driven much of the progress in the field, including many of the contributions celebrated in this issue.

    A natural next step is to test whether vertex-corrected higher-order expansions\cite{kim_pseudoparticle_2022} can restore the correct decoupled-orbital structure at practical cost.
    Additionally, other controlled impurity solvers provide complementary benchmarks \cite{Profumo_Quantum_2015, Bertrand_Quantum_2019, Bertrand_Reconstructing_2019, Macek_Quantum_2020, Bertrand_Quantum_2020,thoenniss_efficient_2023, ng_real-time_2023} Such methods may be useful anchors when leveraging this work towards better methods in future studies.

\section*{Acknowledgments}
    We thank Thomas Blommel and Sergei Iskakov for helpful discussions.
    L.Z.\ and E.G.\ were supported by the National Science Foundation under Grant No. NSF QIS 2310182 and Y.Y.\ by the National Science Foundation under Grant No.\ NSF DMR 2401159.
    G.C.\ acknowledges support from the Israel Science Foundation (Grant No. 2902/21), the PAZY Foundation (Grant No. 318/78) and the MOST NSF-BSF (Grant No. 2023720).
    \\

\bibliography{bib}

\end{document}